\documentclass[hyper,11pt,paper]{article}
\usepackage{jheppub}
\usepackage[usenames,dvipsnames,table]{xcolor}
\usepackage{graphicx,amsmath,amssymb,multirow,array,bm,mathrsfs}
\usepackage{epsf,amsfonts}
\usepackage[numbers,sort&compress]{natbib}

\newcommand{\beq}{\begin{equation}}
\newcommand{\eeq}{\end{equation}}

\newcommand{\tr}{\text{tr}}

\newcommand{\ab}[2]{^{#1}_{\phantom{#1}#2}}

\newcommand{\spp}{L}
\newcommand{\RCP}{R_{{}_\mathbb{CP}}}
\newcommand{\fRCP}{\form{R}_{{}_\mathbb{CP}}}

\newcommand{\half }{\frac{1}{2}}

\newcommand{\form}[1]{\bm{#1}}

\newcommand{\hodge}{{}^\star}

\newcommand{\fA}{\form{A}}

\newcommand{\fF}{\form{F}}

\newcommand{\fB}{\form{B}}

\newcommand{\fGamma}{\form{\Gamma}}

\newcommand{\fR}{\form{R}}

\newcommand{\fBR}{\form{B}_R}

\newcommand{\muR}{{\mbox{\large$\mu$}_R}}

\newcommand{\fu}{\form{u}}
\newcommand{\acc}{a}
\newcommand{\fa}{\form{\acc}}
\newcommand{\fomega}{\form{\omega}}

\newcommand{\fP}{{\form{\mathcal{P}}}}

\newcommand{\VP}{{\form{V}}_{\fP}}

\title{
Chern-Simons terms from thermal circles and anomalies
}
\author[a,b]{Kristan Jensen,}
\author[c]{R. Loganayagam}
\author[d]{Amos Yarom}

\affiliation[a]{Department of Physics and Astronomy, University of Victoria, Victoria, BC V8W 3P6, Canada}
\affiliation[b]{C. N. Yang Institute for Theoretical Physics, SUNY, Stony Brook, NY 11794-3840, United States}
\affiliation[c]{Junior Fellow, Harvard Society of Fellows, Harvard University, Cambridge, MA 02138.}
\affiliation[d]{Department of Physics, Technion, Haifa 32000, Israel}

\emailAdd{kristanj@insti.physics.sunysb.edu}
\emailAdd{nayagam@gmail.com}
\emailAdd{ayarom@physics.technion.ac.il}
\abstract{
We compute the full contribution of flavor and (or) Lorentz anomalies to the thermodynamic partition function. Apart from the Wess-Zumino consistency condition the Euclidean generating function must satisfy an extra requirement which we refer to as `consistency with the Euclidean vacuum.' The latter requirement fixes all Chern-Simons terms that arise in a particular Kaluza-Klein reduction of the theory. The solution to both conditions may be encoded in a `thermal anomaly polynomial' which we compute. Our construction fixes all the thermodynamic response parameters of a hydrodynamic theory associated with anomalies.}
\preprint{YITP-SB-13-38}
\begin{document}
\maketitle

\section{Summary} \label{S:intro}

Anomalies are a ubiquitous feature of quantum field theory which have both experimental and theoretical ramifications. While there is a vast literature concerned with the physical consequences of anomalies in the vacuum state and their cohomological structure, little is known about the manifestation of anomalies in more general states, including thermal states. Recently \cite{Jensen:2012kj,Golkar:2012kb}, following \cite{Bhattacharyya:2007vs,Erdmenger:2008rm,Banerjee:2008th,Torabian:2009qk,Son:2009tf,Kharzeev:2009p,Lublinsky:2009wr,Neiman:2010zi,Bhattacharya:2011tra,Kharzeev:2011ds,Loganayagam:2011mu,Neiman:2011mj,Dubovsky:2011sk,Kimura:2011ef,Lin:2011aa,Loganayagam:2012pz,Gao:2012ix,Banerjee:2012iz,Jensen:2012jy,Jain:2012rh,Valle:2012em,Banerjee:2012cr} (see also~\cite{Chen:2012ca,Manes:2012hf,Loganayagam:2012zg,Bhattacharyya:2013ida,Valle:2013aia} for related works since) it was argued that anomalies lead to distinctive physical phenomena which are only visible near thermal equilibrium. In particular, it follows from the work of \cite{Jensen:2012kj,Golkar:2012kb} that a mixed flavor-gravitational anomaly leads to currents which orient themselves along vorticity, which are in principle measurable in astrophysical phenomena or in condensed matter systems whose low-energy description possesses relativistic fermions. This implies the exciting prospect that mixed anomalies may be measured in Nature.  (See \cite{Chernodub:2013kya} for a recent explicit proposal.)  

In the literature there are various notions of anomalies. Here we focus on anomalies which are shared between global symmetries. That is, the anomalous currents obey non-conservation laws, where the non-conservation is a local function of external sources. It is the anomalies of this type that are exact and must be matched a la `t Hooft across scales. The $U(N_f)_A$ axial symmetry of $N_f$ free, massless Dirac fermions falls into this category, while the $U(1)_A$ axial symmetry of QED does not. In what follows we give a complete classification of the role of anomalies in 
equilibrium configurations. Our results are exact for global anomalies, and only hold perturbatively for the anomalies that involve global and gauge symmetries~\cite{Jensen:2013vta} (see also~\cite{Golkar:2012kb,Hou:2012xg}). 

The theories we study possess global symmetry currents. In what follows we will refer to these currents as flavor currents and to the global symmetry as a flavor symmetry. The theories are coupled to a background gauge field $A_{\mu}$ and a metric $g_{\mu\nu}$. We label the generating functional of the theory as $W_{QFT}[A,g]$. When the theories are anomalous, $W_{QFT}$ is not gauge and coordinate-reparametrization invariant: it varies under an infinitesimal gauge and coordinate variation $\delta_{\chi}$,
\beq
\delta_{\chi}W_{QFT} = \int d^{2n}x \sqrt{-g}\,  G_{\chi}\,,
\eeq
where $G_{\chi}$ is a functional of the background fields and the transformation parameters and we take the theory to live in $2n$ dimensions. The variation $G_{\chi}$ is tightly constrained by the Wess-Zumino (WZ) consistency condition \cite{Wess:1971yu}. In odd dimensions, any local $G_{\chi}$ may be compensated for by a suitable local redefinition of $W_{QFT}$. In contrast, in even space-time dimensions there are local $G_{\chi}$'s which cannot be removed by a local redefinition of $W_{QFT}$.  In what follows, we denote the non gauge invariant contribution to $W_{QFT}$ via $W_{anom}$. Thus,
\beq
\label{E:WQFTsplit}
W_{QFT} = W_{gauge-invariant} + W_{anom}\,.
\eeq

The anomalies of a theory which manifest themselves as $G_{\chi}$ are encoded in the anomaly polynomial $\fP$, which is a closed $2n+2$ form built out of the characteristic classes of the background field strength $\fF = \frac{1}{2}F_{\mu\nu}dx^{\mu}\wedge dx^{\nu}$ and Riemann curvature $\fR\ab{\mu}{\nu} =\frac{1}{2}R\ab{\mu}{\nu\rho\sigma}dx^{\rho}\wedge dx^{\sigma}$. Here and in what follows we use boldface characters to denote form fields. To go from $\fP$ to $G_{\chi}$, one may use the anomaly inflow mechanism of Callan and Harvey~\cite{Callan:1984sa} (for a modern review, see e.g., Appendices A-C of~\cite{Jensen:2013kka}).  

We place our theory in a background with a compact symmetry direction. That is, we put the theory on a manifold given by a circle fibered over a base manifold. In Euclidean thermal field theory we may identify the circle with the thermal circle and $W_{QFT}$ with the logarithm of the thermodynamic partition function. Since the background fields do not depend on the symmetry direction, $W_{QFT}$ is essentially a generating functional on the $2n-1$-dimensional base. (One could imagine compactifying the underlying theory on the circle.) Thus, in an even dimensional theory, whether anomalous or not, $W_{QFT}$ may be reduced to the generating function on the odd-dimensional base manifold. Thus, we expect that in such backgrounds one may find a local expression for $W_{anom}$. In~\cite{Jensen:2013kka} we showed that this is indeed the case by explicitly constructing a local expression for $W_{anom}$.

One might have thought that $W_{gauge-invariant}$ is independent of the anomalies of the theory. This is not the case. There are Chern-Simons (CS) terms on the base manifold which contribute to $W_{gauge-invariant}$. The coefficients of these CS terms are fixed by the coefficients of the anomaly polynomial up to factors of $2\pi$. In what follows we will refer to these CS terms as transcendental terms and denote their contribution to the generating function by $W_{trans}$,
\begin{equation}
	 W_{gauge-invariant} = W_{trans}+W_{non-anomalous}\,.
\end{equation}
The main goal of this work is to determine the CS coefficients in $W_{trans}$ and their relation to the anomalies.

Calculations for free Weyl fermions in four~\cite{Vilenkin:1980fu,Landsteiner:2011cp} and arbitrary dimension~\cite{Loganayagam:2012pz} have shown that these CS coefficients (more precisely, even the CS terms on the base manifold which do not involve the gravitational connection) are proportional to gravitational anomaly coefficients.  Computations carried out in the framework of the AdS/CFT correspondence for four dimensional theories give similar results \cite{Landsteiner:2011iq}. Recently, these computations have been extended to arbitrary dimension \cite{RRHarvard}. This has led to a conjecture that the coefficients of the CS terms on the base manifold are determined by the anomalies in an arbitrary interacting theory. This conjecture was verified in two~\cite{Jensen:2012kj} and four~\cite{Jensen:2012kj,Golkar:2012kb} dimensions. In this work, we extend the technique of~\cite{Jensen:2012kj} to verify this conjecture and go beyond it. Let us review the thrust of the argument of~\cite{Jensen:2012kj} for two dimensional theories. Consider a two dimensional field theory with a gravitational anomaly placed on $\mathbb{R}^{2,*}$, the Euclidean plane with the origin removed and trivial boundary conditions have been imposed there. We identify the angular direction on  $\mathbb{R}^{2,*}$ with Euclidean time. The thermal partition function must then reproduce all rotationally-invariant correlation functions of the Euclidean vacuum on $\mathbb{R}^2$, including the one-point function of the stress tensor.\footnote{Upon Wick-rotation, this spacetime becomes a Rindler wedge. Our claim is tantamount to the statement that the thermal partition function on Rindler space at temperature $1/(2\pi)$ computes boost-invariant correlation functions for operator insertions within the wedge, including the energy-momentum tensor. See e.g.~\cite{Eling:2013bj} for related discussion.} 
In what follows we will refer to this property as ``consistency with the Euclidean vacuum,'' and impose it as a consistency requirement on $W_{QFT}$. This requirement is non-trivial and relates the gravitational anomaly of the two-dimensional theory to a Chern-Simons coefficient on the base manifold.

Imposing consistency with the Euclidean vacuum, we find that the physics of anomalies in the thermodynamic partition function may be encoded in a ``thermal anomaly polynomial'' $\fP_T$. The thermal anomaly polynomial is obtained from the anomaly polynomial $\fP$ via an algorithm, the ``replacement rule.''
It amounts to the following statement. Consider an anomaly polynomial $\fP$, which we may view as a function of the Chern classes of $\fF$ and the Pontryagin classes $\form{p}_k(\fR)$ of $\fR\ab{\mu}{\nu}$ (see \eqref{E:defpk} for a concise definition of Pontryagin classes). In terms of these we define $\fP_T$ as
\beq
\label{E:PT1V2}
\fP_T = \fP\left(\fF,\form{p}_k(\fR)\to \form{p}_k(\fR)-\left( \frac{\fF_T}{2\pi}\right)^2 \wedge \form{p}_{k-1}(\fR)\right)\,,
\eeq
where we have introduced a spurious $U(1)$ gauge symmetry with connection $\fA_T$ and field strength $\fF_T=d\fA_T$ whose role will become clear shortly. We work in a convention where $\form{p}_0(\form{R})=1$.

A precursor to the ``replacement rule'' was conjectured in \cite{Loganayagam:2012pz} using a slightly different formalism than ours which involves thermal helicity correlators~\cite{Loganayagam:2012zg}. The conjecture of \cite{Loganayagam:2012pz} was based on results for free Weyl fermions \cite{Vilenkin:1980fu,Landsteiner:2011cp,Loganayagam:2012pz} and has been recently found to hold in the context of holography~\cite{RRHarvard}. A concise statement which we can make is that the conjecture of~\cite{Loganayagam:2012zg}, in the current language, is an assertion about the part of $\fP_T$ in~\eqref{E:PT1V2} that sends $\form{p}_1(\fR)\to -\left(\frac{\fF_T}{2\pi}\right)^2$. The present work extends the conjecture of~\cite{Loganayagam:2012zg}, and proves it.

Using $\fP_T$ we construct a master function $\form{V}_T$ which is a $2n+1$ form whose derivatives give us the entire contribution of the anomaly to the flavor current and energy-momentum tensor. To explain the construction of $\form{V}_T$ we need to use the background fields and symmetry data to construct a number of useful quantities.

We collectively denote the symmetry data as $K=\{K^{\mu},\Lambda_K\}$, where $K^{\mu}$ is a timelike vector and $\Lambda_K$ is a gauge transformation parameter. We label the corresponding variation by $\delta_K$. When we say that $K$ generates a symmetry, we mean that $\delta_K$ vanishes when acting on the background. From $\{K^{\mu},\Lambda_K\}$ we may define a local temperature, velocity field, and flavor chemical potential via
\beq
\label{E:hydroVarDef}
T \equiv \frac{1}{\beta \sqrt{-K^2}}\,, \qquad u^{\mu} \equiv \frac{K^{\mu}}{\sqrt{-K^2}}\,, \qquad \mu \equiv \frac{K^{\alpha}A_{\alpha}+\Lambda_K}{\sqrt{-K^2}}\,.
\eeq
The parameter $\beta$ specifies the affine periodicity of the thermal circle (the integral curves of $K^{\mu}$). The definitions~\eqref{E:hydroVarDef} are constructed to coincide with the standard temperature, velocity, and chemical potential in the source-free thermal state (e.g., $T$ corresponds to the inverse length of the thermal circle). As explained in detail in~\cite{Jensen:2013kka}, from these variables one can construct the spin chemical potential
\beq
(\mu_R)\ab{\mu}{\nu} \equiv T D_{\nu}\left( \frac{u^{\mu}}{T}\right)\,.
\eeq

Using the velocity one-form $\form{u} = u_{\mu}dx^{\mu}$ one may construct hatted connections,
\beq
\label{E:hatDef}
\hat{\fA} \equiv \fA + \mu \fu\,, \qquad \hat{\fGamma}\ab{\mu}{\nu} \equiv \fGamma\ab{\mu}{\nu}+(\mu_R)\ab{\mu}{\nu}\fu\,, \qquad \hat{\fA}_T \equiv \fA_T + \mu_T \fu\,,
\eeq
where $\fGamma\ab{\mu}{\nu}=\Gamma\ab{\mu}{\nu\rho}dx^{\rho}$ is the Christoffel connection one-form, and  we have defined
\beq
\mu_T \equiv 2 \pi T\,.
\eeq
One may also define the corresponding hatted field strengths $\hat{\fF}, \hat{\fR}\ab{\mu}{\nu}$, and $\hat{\fF}_T$. The vorticity $2\fomega$, the magnetic flavor field $\fB$, and the magnetic curvature $(\fB_R)\ab{\mu}{\nu}$ are defined via
\beq
\omega_{\mu\nu} \equiv \frac{\partial_{\rho}u_{\sigma} - \partial_{\sigma}u_{\rho}}{2}P\ab{\rho}{\mu}P\ab{\sigma}{\nu}\,, \qquad B_{\mu\nu} \equiv F_{\rho\sigma}P\ab{\rho}{\mu}P\ab{\sigma}{\nu}\,, \qquad (B_R)\ab{\mu}{\nu\rho\sigma}\equiv R\ab{\mu}{\nu\alpha\beta}P\ab{\alpha}{\rho}P\ab{\beta}{\sigma}\,,
\eeq
for $P_{\mu\nu}\equiv g_{\mu\nu}+u_{\mu}u_{\nu}$ the transverse projector. The magnetic component associated with the spurious $U(1)$ symmetry, $\form{B}_T$, is similarly defined.

The master function $\form{V}_T$ is given by
\beq
\label{E:VTdef}
\form{V}_T \equiv \frac{\fu}{2\fomega}\wedge \left( \fP_T -\hat{\fP}_T\right)\,,
\eeq
where by $\hat{\fP}_T$ we mean $\fP_T(\hat{\fF},\hat{\fR},\hat{\fF}_T)$. The inverse factor of $2\fomega$ is a shorthand for the following. The $2n+3$ form $\fu\wedge \left( \fP_T-\hat{\fP}_T\right)$ is a polynomial of at least degree 1 in $(2\fomega)$, $\sum_{k=1}^{n+1} \form{c}_k \wedge (2\fomega)^k$ for $\form{c}_k$ a $2n-2k+3$ form. The division by $2\fomega$ is an instruction to remove a power of $(2\omega)$ to give $\sum_{k=0}^n \form{c}_{k+1} \wedge (2\fomega)^k$. If we regard $\form{V}_T$ as a functional whose independent variables are the velocity $\fu$, the chemical potentials $\{\mu,\mu_R,\mu_T\}$, the magnetic fields $\{\fB,(\fB_R)\ab{\mu}{\nu}\,\form{B}_T\}$, and the vorticity $(2\fomega)$, then the (Hodge duals of) the anomaly-induced flavor, heat, and spin currents are (see Appendix I of~\cite{Jensen:2013kka} for our conventions for the Hodge star)
\beq
\label{E:Ttransport}
\hodge \form{J}_T = \frac{\partial \form{V}_T}{\partial \fB}\,, \qquad \hodge \form{q}_T = \frac{\partial \form{V}_T }{\partial (2\fomega)}\,, \qquad \hodge (\form{\spp}_T)\ab{\mu}{\nu}= \frac{\partial \form{V}_T}{\partial (\fB_R)\ab{\nu}{\mu}}\,,
\eeq
all evaluated at $\fF_T=0$. The anomaly-induced stress tensor is given by the combination
\beq
\label{E:TT}
T_T^{\mu\nu} = u^{\mu}q^{\nu}+u^{\nu}q^{\mu} + D_{\rho}\left[ \spp_T^{\mu[\nu\rho]}+\spp_T^{\nu[\mu\rho]} - \spp_T^{\rho(\mu\nu)}\right]\,,
\eeq
where $D_{\mu}$ is the covariant derivative and the brackets indicate (anti)symmetrization
\beq
A^{(\mu\nu)} = \frac{1}{2}(A^{\mu\nu}+A^{\nu\mu})\,, \qquad A^{[\mu\nu]} = \frac{1}{2}(A^{\mu\nu}-A^{\nu\mu})\,.
\eeq
Using the same formalism, we also obtain an explicit expression for the contribution of the anomalies to $W_{QFT}$. We find that, in a gauge and coordinate choice where $\{A_{\mu},g_{\mu\nu}\}$ are explicitly time-independent,
\begin{equation}
\label{E:Wtrans}
	W_{trans}+W_{anom} = -\int \frac{\form{u}}{2\form{\omega}} \wedge \left(\form{I}_T - \hat{\form{I}}_T\right)_{\fA_T=0,\fF_T=0}\,,
\end{equation}
where $\form{I}_T$ is the Chern-Simons term associated with $\fP_T$, i.e., $d\form{I}_T = \fP_T$. Similarly, $d\hat{\form{I}}_T = \hat{\fP}_T$.

As we have already mentioned, our construction relies on a certain consistency condition with the Euclidean vacuum. We believe that this is not the most elegant way to obtain our results. We say this on moral grounds---the argument we have used strikes us as unnecessarily coarse given the mathematical elegance of anomalies---but we also have some recent results in mind. As explained in \cite{Jensen:2012kj} the consistency condition with the Euclidean vacuum breaks down in the presence of gravitinos. An explicit computation~\cite{Loganayagam:2012zg} for weakly coupled theories of chiral gravitinos has shown that the partition function does not take the form  \eqref{E:Wtrans}. Perhaps relatedly, one may expect that coefficients of CS terms satisfy a subtle quantization condition.\footnote{We thank Z.~Komargodski and D.~Son for discussions on this point.} Thus, it seems that a cleaner argument for fixing $W_{trans}$ should exist.

That being said, our results are interesting on several levels. One area where they have physical implications is in the hydrodynamic limit of anomalous field theories. In what follows, we discuss this relation, but we emphasize that the results above stand on their own without reference to hydrodynamics.

Hydrodynamics may be thought of as the low-energy effective description of thermal field theory. Its degrees of freedom correspond to the conserved charges, and may be chosen to be a local temperature $T$, a local chemical potential $\mu$ and a local velocity field $u^{\mu}$ normalized such that $u^{\mu}u_{\mu}=-1$. The stress tensor and flavor currents may be thought of as functions of the hydrodynamic variables and slowly varying background fields in a derivative expansion \cite{Baier:2007ix,Bhattacharyya:2008jc}. The resulting expansion of the stress tensor and current are referred to as the constitutive relations. In a Lorentz-invariant theory the constitutive relations are fixed up to some scalar coefficients (e.g., the conductivity or shear viscosity) which we term response parameters. The hydrodynamic variables $\{T,\mu,u^{\mu}\}$ are then determined by demanding that the stress tensor and current solve the corresponding Ward identities, which are regarded as equations of motion.

The response parameters of hydrodynamics are constrained by an internal consistency condition, which amounts to a local version of the second law of thermodynamics. One requires the existence of an entropy current $s^{\mu}$, whose divergence is non-negative for solutions of the hydrodynamic equations~\cite{landau_fluid_1987}. This requirement is surprisingly restrictive, and among other things it fixes equality-type and inequality-type interrelations between various response parameters.

One may use a thermal partition function to compute zero-frequency correlation functions at low-momentum. To do so, one assumes a finite static screening length, in which case the partition function may be expanded in a derivative expansion~\cite{Banerjee:2012iz,Jensen:2012jh}. The consequent correlation functions are also ostensibly computed by hydrodynamics, or more precisely, hydrostatics. Upon matching the two, one finds that, for all cases discussed in the literature so far, the thermodynamic partition function precisely reproduces the equality-type interrelations demanded by the existence of an entropy current. This matching has led to a conjecture~\cite{Banerjee:2012iz,Jensen:2012jh} that the equality type relations of hydrodynamics associated with response parameters are fully reproduced by the thermal partition function. If this conjecture is true, then our analysis gives us the complete set of thermodynamic response parameters which are completely fixed by the anomaly via equality type relations. In a companion paper~\cite{companion} we show that any thermodynamic partition function, including the contributions from anomalies, is consistent with the existence of an entropy current. In particular, we compute a (Hodge dual of a) representative of the contribution of the anomaly to the entropy current, ${S}_{T}^{\mu}$, which is given by 
\beq
\hodge \form{S}_T = 2\pi \frac{\partial \form{V}_T}{\partial \fB_T}\,,
\eeq
evaluated at $\fF_T=0$, plus terms that vanish in hydrostatic equilibrium.

The rest of this manuscript is organized as follows. In Section~\ref{S:HydroCS} we review the results of~\cite{Jensen:2013kka} required for the rest of our work including, in particular, the construction of $W_{anom}$ and the currents derived from it. We also describe how one can repackage CS terms on the $2n-1$ dimensional base manifold as $2n+1$ dimensional form fields. In Section~\ref{S:VT} we demand consistency with the Euclidean vacuum and thereby fix the Chern-Simons coefficients of the thermodynamic partition function. In the same Section we also demonstrate the various claims in this Summary, including the form of the thermal anomaly polynomial and $\form{V}_T$. We provide a detailed exposition of our results in Appendix~\ref{A:examples} for two, four, and ten dimensional theories. 

\section{Components of the generating functional}
\label{S:HydroCS}

Consider a quantum field theory coupled to background fields which posses a timelike symmetry. We will refer to the partition function of the theory in this background, evaluated when the time direction has been Wick-rotated to Euclidean signature and compactified, as the thermodynamic partition function $Z_E$. It is related to the generating functional $W_{QFT}$ by $W_{QFT} = - i \ln Z_E$. We refer to the resulting state as an equilibrium state, bearing in mind that it has spatial gradients. 

A hydrostatic configuration is an example of an equilibrium state in which no entropy is generated and where the background fields are slowly varying.
A column of air in the atmosphere is an example of such a state. By varying the thermodynamic partition function with respect to the background fields we obtain correlation functions of the theory in a hydrostatic configuration. These correlation functions, including the one-point functions of the stress tensor and symmetry currents, allow us to relate the thermodynamic partition function to hydrodynamics~\cite{Banerjee:2012iz,Jensen:2012jh}.

As described in detail in \cite{Jensen:2013kka}, if we denote the generators of the timelike symmetry by $\{K^{\mu},\Lambda_K\}$ (where $\Lambda_K$ is a gauge transformation parameter), then in equilibrium the temperature, velocity field and chemical potential are given by \eqref{E:hydroVarDef}. In what follows we will work in a particularly useful gauge, the ``transverse gauge,'' where we take the background to be explicitly time-independent. This amounts to taking $\Lambda_K=0$ and  $K^{\mu}\partial_{\mu} = \beta \partial_t$ for $\beta$ the parametric length of the Euclidean time circle. The metric and gauge field may be written in the form
\begin{align}
\begin{split}
\label{E:transversedef}
	g &= -e^{2\mathfrak{s}(x)}(dt + \mathfrak{a}_i(x) dx^i)^2+\mathfrak{p}_{ij}(x)dx^i dx^j \,,\\
	\form{A} &= A_0(x)(dt+\mathfrak{a}_i(x) dx^i) + \mathfrak{A}_i(x) dx^i 
\end{split}
\end{align}
in which case the relations \eqref{E:hydroVarDef} and~\eqref{E:hatDef} reduce to
\begin{equation}
\label{E:HydroVarTransverse}
	\beta T = e^{-\mathfrak{s}}\,,
	\qquad
	\form{u} = -e^{\mathfrak{s}}(dt + \form{\mathfrak{a}})\,,
	\qquad
	\frac{\mu}{T} = \beta A_0\,,
	\qquad
	\hat{\fA} = \mathfrak{A}_idx^i\,.
\end{equation}
We remind the reader that we consistently use boldface characters for form fields. We also refer the reader to~\cite{Banerjee:2012iz} for a thorough discussion of the transverse gauge and to~\cite{Jensen:2013kka} for a covariant description of equilibrium states.

Our results crucially rely on the properties of equilibrium states, as well as on the machinery developed in~\cite{Jensen:2013kka}. In what follows we briefly review the salient features of~\cite{Jensen:2013kka} we require for our analysis, and go on to study the Chern-Simons terms on the spatial slice thereafter.

\subsection{Review of $W_{anom}$}

In the presence of anomalies the generating functional $W_{QFT}$ must satisfy the Wess-Zumino consistency conditions. As described in Section \ref{S:intro}, one may split $W_{QFT}$ into a gauge-invariant contribution $W_{gauge-invariant}$ and a non gauge-invariant and (or) non diffeomorphism invariant contribution $W_{anom}$, see \eqref{E:WQFTsplit}. The Wess-Zumino consistency condition  fixes the allowed variations of $W_{anom}$ under a gauge and (or) coordinate transformation. This variation is determined via the descent relations in terms of an anomaly polynomial $\fP$.  In thermal equilibrium and in transverse gauge one can obtain an explicit local expression for $W_{anom}$: given the anomaly polynomial  one can construct a $2n$ form
\begin{equation}
\label{E:WCS}
	\form{W}_{CS} = \frac{\form{u}}{2\form{\omega}} \wedge \left(\form{I} - \hat{\form{I}}\right)
\end{equation}
where $\form{I}$ is a Chern-Simons term associated with the anomaly polynomial $\fP$ via $d\form{I} = \fP$ and with the conventions described in Section \ref{S:intro} such that $W_{anom} = -\int \form{W}_{CS}$. See~\cite{Jensen:2013kka} for details. Varying $W_{QFT}$ with respect to the background sources will give us the connected correlators in 
an equilibrium state. In particular, varying $W_{anom}$ with respect to the sources will give contributions to the connected correlators which are proportional to the anomaly coefficients. 

The currents obtained by varying the generating functional $W_{QFT}$ are referred to as consistent currents since they are generated by a functional which satisfies the Wess-Zumino consistency condition. The consistent currents have the unfortunate property that they are non-gauge and (or) diffeomorphism covariant \cite{Bardeen:1984pm}. Luckily, one may add to the consistent currents polynomials local in the sources (Bardeen-Zumino polynomials~\cite{Bardeen:1984pm}) which render the total expression covariant. The latter currents are called covariant currents. The generating functional for covariant currents is given by \cite{Callan:1984sa},
\begin{equation}
	W_{cov} = W_{QFT} + \int_{\mathcal{M}} \form{I}
\end{equation}
where $\mathcal{M}$ is a manifold on whose boundary $W_{QFT}$ is defined. After some massaging, one can show that
\begin{equation}
\label{E:Wcov}
	W_{cov} = W_{gauge-invariant}+\int_{\mathcal{M}} \VP
\end{equation}
with
\begin{equation}
	\VP = \frac{\form{u}}{2\form{\omega}} \wedge \left(\fP - \hat{\fP}\right)\,.
\end{equation}
The contribution of $W_{anom}$ to the covariant currents is given by
\begin{equation}
\label{E:currentP}
	\hodge \form{J}_{\fP} = \frac{\partial \VP}{\partial \form{B}}\,, \qquad
	\hodge \form{q}_{\fP} = \frac{\partial \VP}{\partial(2 \form{\omega})}\,, \qquad
	\hodge (\form{L}_{\fP})\ab{\mu}{\nu} = \frac{\partial \VP}{\partial (\form{B}_R)\ab{\nu}{\mu}}\,.
\end{equation}
Here, we have represented the flavor current $J^{\mu}_{\mathcal{P}}$, the heat current $q^{\mu}_\mathcal{P}$, and the spin current $(L_{\mathcal{P}})\ab{\rho\mu}{\nu}$ in terms of their Hodge duals. The heat and spin currents determine the stress tensor via
\begin{equation}
\label{E:Tp}
	T^{\mu\nu}_{\mathcal{P}} = 2 u^{(\mu} q_\mathcal{P}^{\nu)} + D_{\lambda} \left(L_{\mathcal{P}}^{\mu[\nu\lambda]} + L_{\mathcal{P}}^{\nu[\mu\lambda]}-L_{\mathcal{P}}^{\lambda(\mu\nu)}\right)\,.
\end{equation}
We refer the interested reader to \cite{Jensen:2013kka} for a derivation. 

\subsection{The thermal anomaly polynomial and $W_{trans}$}
\label{S:WtransCS}

The subscript $\fP$ in \eqref{E:currentP} has been used to emphasize that these are not the full currents of the theory but only a particular additive contribution to these currents which comes from the non-gauge and (or) diffeomorphism invariant part of $W_{QFT}$. Naively one would think that the remaining gauge-invariant part of $W_{QFT}$ is oblivious to the anomalies. The goal of this paper is to argue that this is not the case. Let us think of the Euclidean space-time on which the theory is defined as a thermal circle fibered over a base manifold. The gauge-invariant components of the generating functional may be split into manifestly gauge and diffeomorphism invariant terms and Chern-Simons terms on the base manifold. We will refer to the component of the generating functional which includes these Chern-Simons terms as $W_{trans}$,
\begin{equation}
	W_{gauge-invariant} = W_{non-anomalous}+W_{trans}\,.
\end{equation}
In this work, we argue that the coefficients of the Chern-Simons terms on the base-manifold are fixed by the anomaly coefficients (as they appear in the anomaly polynomial) up to factors of $\pi$, hence the subscript.

As a first step towards proving this claim, we will recast the expressions for Chern-Simons terms on the base manifold into a form which is more reminiscent of the structure of $W_{anom}$, e.g., equations \eqref{E:WCS} and \eqref{E:Wcov}. To do so, we introduce a fictitious abelian gauge connection $\fA_T$ whose fictitious field strength is $\fF_T=d\fA_T$. Its usefulness will become clear shortly. We define the chemical potential for this fictitious symmetry as
\beq
\mu_T \equiv 2 \pi T\,.
\eeq
We will ultimately set $\fA_T$ to vanish. Now, consider the $2n+2$ form 
\begin{equation}
\label{E:Pcsb}
	 \qquad \fP_{trans} = \sum_{q}{\fF_T}^{q} \wedge \sum_{i_q} c_{i_q} \form{P}_{i_q}\left(\form{F},\,\form{R}\right)
\end{equation}
where $\form{P}_{i_q}$ are various possible exact $2(n+1-q)$ forms constructed out of wedge products of flavor field strength $\fF$ and Riemann curvature $\fR\ab{\mu}{\nu}$. Invariance under CPT implies that $q$ takes on even values. Consider the combinations
\begin{align}
\begin{split}
\label{E:csbpotentials}
	\form{V}_{trans} &= \frac{\form{u}}{2\form{\omega}} \wedge \left(\fP_{trans} - \hat{\fP}_{trans}\right)\Bigg|_{\form{F}_T=0}\,, \\
	\form{W}_{trans} &= \frac{\form{u}}{2\form{\omega}} \wedge \left(\form{I}_{trans} - \hat{\form{I}}_{trans}\right)\Bigg|_{\form{A}_T=0\,, \form{F}_T=0}\,,
\end{split}
\end{align}
where $\fP_{trans} = d\form{I}_{trans}$. We claim that the Chern-Simons forms associated with $W_{trans}$ are captured by \eqref{E:csbpotentials} via\footnote{In defining $\form{W}_{trans}$ as in~\eqref{E:csbpotentials}, we have adopted the notation used in~\cite{Jensen:2013kka}, which has the unfortunate byproduct that $W_{trans} = - \int_{\partial\mathcal{M}}\form{W}_{trans}$. We hope that this will not cause confusion.}
\begin{align}
\begin{split}
\label{E:csbWs}
	W_{cov} &= W_{non-anomalous} + \int_{\mathcal{M}}\left( \form{V}_{\fP} + \form{V}_{trans}\right) \,,\\
	W_{QFT} & = W_{non-anomalous} - \int_{\partial\mathcal{M}} \left( \form{W}_{CS} + \form{W}_{trans}\right) \,.
\end{split}
\end{align}
To prove that \eqref{E:csbWs} is correct, it is sufficient to show that $\form{W}_{trans}$ corresponds to a Chern-Simons term on the base manifold and that $d\form{W}_{trans} = -\form{V}_{trans}$.

Let us consider a particular representative of $\fP_{trans}$, 
\begin{equation}
\label{E:Ptransrep}
	\fP_{trans} = {\fF_T}^q\wedge \form{P}(\form{F},\,\form{R})\,.
\end{equation}
The associated $\form{V}_{trans}$ is given by
\begin{align}
\begin{split}
	\form{V}_{trans} &= \frac{\form{u}}{2\form{\omega}} \wedge \left( \fP_{trans} - \hat{\fP}_{trans} \right) \Big|_{\form{F}_T=0} \\
		& = \frac{\form{u}}{2\form{\omega}} \wedge \left({\form{B}_T}^q\wedge \form{P}(\form{B},\,\form{B}_R) - (\form{B}_T+2 \form{\omega}\mu_T)^q\wedge \form{P}(\hat{\form{B}},\,\hat{\form{B}_R}) \right) \Big|_{\form{F}_T=0} \\
		& = -\mu_T\form{u} \wedge (2 \form{\omega}\mu_T)^{q-1} \wedge \hat{\form{P}} \,.
\end{split}
\end{align}
Here we and are working in our standard convention where hatted forms are evaluated with hatted connections so that $\hat{\fB}_T = \fB_T+2\fomega\mu_T$. 

Using $\form{P} = d\form{i}$ we take $\form{I}_{trans}$ in~\eqref{E:csbpotentials} to be $ \form{I}_{trans} = \fF_T^q \wedge \form{i}$ so that
\begin{align}
\begin{split}
	\form{W}_{trans} & = \frac{\form{u}}{2\form{\omega}} \wedge \left( {\form{B}_T}^q\wedge  \form{i} - (\form{B}_T+2 \form{\omega}\mu_T)^q\wedge  \hat{\form{i}} \right)\Big|_{\fA_T=0\,,\form{F}_T=0}  \\
	&= -\mu_T \form{u} \wedge (2\form{\omega}\mu_T)^{q-1}\wedge \hat{\form{i}} \,.
\end{split}
\end{align}
We now go to the particular decomposition associated with the transverse gauge given in \eqref{E:HydroVarTransverse} under which $\mu_T\form{\omega} =- \frac{2\pi}{\beta} d\form{\mathfrak{a}}$ and $\mu_T \form{u} = -\frac{2\pi}{\beta}(dt+\form{\mathfrak{a}})$. We find that
\begin{equation}
	\int_{\mathcal{M}}\form{V}_{trans} = \int_{\partial \mathcal{M}}\form{W}_{trans}
\end{equation}
where
\begin{equation}
\label{E:W0isCSB}
	\int_{\partial \mathcal{M}}\form{W}_{trans} =\int_{\partial\mathcal{M}} \left(-\frac{2\pi}{\beta}\right)^q dt \wedge (d\form{\mathfrak{a}})^{q-1} \wedge \form{i}\left(\hat{\form{A}},\hat{\form{B}},\hat{\form{\Gamma}},\,\hat{\form{B}}_R\right)\,.
\end{equation}
Thus, we find that, after integrating over the time circle, $\int \form{W}_{trans}$ reduces to a Chern-Simons form on the base manifold. Moreover, any Chern-Simons term on the base manifold can be written in the form \eqref{E:W0isCSB} after integrating by parts.

We can now state the goal of this paper more concisely. We claim that all the coefficients of the Chern-Simons terms on the base manifold are fixed in terms of the coefficients of the anomaly polynomial. More succinctly, we claim that
\begin{equation}
\label{E:RRv1}
	\fP_T = \fP+\fP_{trans}
\end{equation}
where $\fP_T$ is determined via the replacement rule \eqref{E:PT1V2} and the  parameters which are determined by this equation are the coefficients $c_{i_q}$ of \eqref{E:Pcsb}. In the next Section we will use a novel consistency condition to argue for \eqref{E:RRv1}.

\section{Obtaining the master function $\mathbf{V}_T$}
\label{S:VT}
We now turn to our main argument which fixes the coefficients of $W_{trans}$ (or alternately $\form{V}_T$) via the replacement rule \eqref{E:PT1V2}. Our argument is essentially a generalization of the technique used in~\cite{Jensen:2012kj} where we derived $W_{trans}$ for two and four dimensional theories. Our procedure for computing $W_{trans}$ is as follows. We use $W_{cov}$ to compute a particular correlator that should vanish in the Euclidean vacuum. This correlator is described in Subsection~\ref{S:observables}. In Subsection~\ref{S:perturbative} we argue that $W_{non-anomalous}$ does not contribute to such a correlator. We then find in Subsection~\ref{S:RR} that this correlation function does not vanish unless the replacement rule is satisfied.

\subsection{The setup}
\label{S:observables}

We place our theory in a highly symmetric $2n$ dimensional background $\mathcal{N}$ given by
\beq
\label{E:Ndef}
\mathcal{N} = \mathbb{R}^{1,1}\times\mathbb{R}^{2k}\times\mathbb{R}^{4l} \times (\mathbb{CP}^{2m_1} \times \hdots\times  \mathbb{CP}^{2m_p})\,,
\eeq
with an infinitesimally small but covariantly constant flavor magnetic flux threading the $\mathbb{R}^{2k}$ plane, and give the $\mathbb{R}^{4k}$ an infinitesimal angular velocity corresponding to a Kaluza Klein flux along the thermal circle in the Euclidean version of the space-time. 

To be more precise, let us label the coordinates along $\mathbb{R}^{2k}$ as $x^a$, the coordinates along $\mathbb{R}^{4l}$ as $y^i$ and the coordinates on the product of $\mathbb{CP}^m$ spaces by $z^{\alpha}$. We span the Euclidean version of $\mathbb{R}^{1,1}$ with polar coordinates such that they cover the punctured plane $\mathbb{R}^{2,*}$ where the time coordinate is identified with the angular direction. That is, in Lorentzian signature we use an $r$, $t$ coordinate system which covers a Rindler wedge of $\mathbb{R}^{1,1}$. 

The infinitesimally small, but covariantly constant, flavor magnetic fields we turn on arise from a flavor connection
\beq
\label{E:RRgauge}
\fA = \frac{1}{2}B_{ab} x^a dx^b\,,
\eeq
where $B_{ab}=-B_{ba}$ are constant matrices valued in the Cartan subalgebra of the symmetry algebra $\mathfrak{g}$ so that $[B_{ab},B_{cd}]=0$. The resulting flavor field strength is 
\beq
\label{E:RRF}
\fF = \frac{1}{2}B_{ab} dx^a\wedge dx^b\,.
\eeq
By construction $D_{\mu}F_{\nu\rho} = 0$. We thread the entire $\mathbb{R}^{2k}$ plane with flavor magnetic flux. By a suitable $SO(2k)$ transformation, we redefine the $x^a$ so that the only nonzero components of $B_{ab}$ are $\{B_{12}, B_{34}, \hdots, B_{(2k-1)2k}\}$. We then work in a perturbative expansion in which we neglect terms which include two powers of the same $B_{ab}$. Within this perturbative scheme, we can turn on any nonzero Chern class of $\fF$ on the $\mathbb{R}^{2k}$ plane.

For the background metric we use the coordinate system
\beq
\label{E:RRmetric}
g = -r^2 \left( dt + \frac{1}{2}b_{ij}y^i dy^j\right)^2 + dr^2 + \delta_{ab}dx^adx^b+\delta_{ij}dy^idy^j + G_{\alpha\beta}dz^{\alpha}dz^{\beta}\,,
\eeq
where $b_{ij}=-b_{ji}$ are constants and $G_{\alpha\beta}dz^{\alpha}dz^{\beta}$ denotes the Fubini-Study metric on the $\mathbb{CP}^{2m_1}\times\hdots \times \mathbb{CP}^{2m_p}$ spaces. (Here we have suppressed radii of curvature $R_{m_i}$ for each $\mathbb{CP}^{2m_i}$ space, as these radii decouple from our analysis.) Note that the Riemann curvature constructed from the Fubini-Study metric is covariantly constant,
\beq
D_{\mu}(\RCP)\ab{\nu}{\rho\sigma\tau} = 0\,,
\eeq
similar to the constant flavor field strengths we turned on. The resulting Riemann curvature threads the $z$ directions. By a suitable choice of $m_i$, we can turn on any Pontryagin class (with index smaller than $\sum_i m_i$) of $\fR\ab{\alpha}{\beta}$ in the $z$ directions, in the same way that we were able to turn on any Chern class of $\fF$  (with index less than $k$) on the $\mathbb{R}^{2k}$ plane.

The metric \eqref{E:RRmetric} is written in a transverse gauge as in~\eqref{E:transversedef}, with a Kaluza-Klein connection
\beq
\form{\mathfrak{a}} = \frac{1}{2}b_{ij}y^i dy^j\,,
\eeq
from which we find a Kaluza-Klein field strength
\beq
\form{\mathfrak{f}} = d\form{\mathfrak{a}} = \frac{1}{2}b_{ij}dy^i \wedge dy^j\,,
\eeq
i.e. we have turned on constant KK magnetic fields in the $y$ directions. Paralleling our discussion of flavor magnetic fields, we choose the $b_{ij}$ so that the KK flux threads the $\mathbb{R}^{4l}$ plane, or equivalently $d\form{\mathfrak{a}}^{2l}\neq 0$. We also perform an $SO(4l)$ transformation to rotate the $y$'s so that the only nonzero components of $b_{ij}$ are $\{b_{12},b_{34},\hdots b_{(4l-1)4l}\}$, and moreover work in a perturbative expansion wherein we neglect more than one power of the same $b_{ij}$. In this limit, the KK field strength is covariantly constant, $D_{\mu} f_{jk} = 0$. 

Correlation functions obtained from the thermodynamic partition function, evaluated on the backgrounds \eqref{E:Ndef} must agree with Euclidean vacuum correlators on $\mathbb{R}^{2+2k+4l}\times\left(\mathbb{CP}^{2m_1}\times\hdots \times \mathbb{CP}^{2m_p}\right)$. More precisely, they must agree with zero frequency vacuum correlators (where the frequency is the conjugate to Rindler time $t$). We term this property, which we view as a consistency condition on the Euclidean generating functional, ``consistency with the Euclidean vacuum.'' 

In the remainder of this Section we will check consistency of $\langle T_{cov}^{tr} \rangle$ with the Euclidean vacuum. We will show that for generic coefficients of $W_{trans}$, $\langle T_{cov}^{tr} \rangle$  will be proportional to $\text{det}\, B\,\text{det}\,b$.
By taking appropriate derivatives with respect to the external flavor gauge field and Kaluza-Klein field we can obtain from  $\langle T_{cov}^{tr} \rangle$ a connected correlator in the Euclidian vacuum of $\mathcal{N}$ in the absence of Kaluza-Klein and magnetic fields. We will refer to such a correlator as $C$. 
More formally, the correlator $C$ has $k$ current insertions, $J^a$, and $2l$ energy flux  insertions, $T^{0i}$. The currents carry momenta in the $\mathbb{R}^{2k}$ plane which are orthogonal to each other and to the current insertions. Similarly, the momenta carried by the energy flux insertions $T^{0i}$ are also orthogonal to each other and to the energy flux directions. That is, the correlator $C$ obtained from $\langle T_{cov}^{tr}\rangle$ is
\beq
	C = \langle T^{tr}_{cov} T^{0 i_1}(p_1) \ldots T^{0i_{2l}}(p_{2l}) J^{a_1}(q_1) \ldots J^{a_k}(q_k)\rangle\,,
\eeq
where the $T_{cov}^{tr}$ insertion carries momentum $- \sum_i p_i - \sum_a q_a$. Due to the product structure of $\mathcal{N}$ and the fact that $C$ is a scalar from the point of view of the $\mathbb{CP}^m$, $C$ is of the form
\beq
\label{E:Cdef}
C \sim  \epsilon^{i_1\hdots i_{2l}a_1\hdots a_k} (p_1)_{i_1}\hdots (p_{2l})_{i_{2l}} (q_1)_{a_1}\hdots (q_k)_{a_k} \,.
\eeq
The $\epsilon$ tensor in \eqref{E:Cdef} is the epsilon tensor on the $\mathbb{R}^{2k+4l}$ space.

The objects available to us for constructing the correlator $C$ in the Euclidean vacuum are the epsilon tensor, the metric, and the various momenta. As we now show, these are insufficient to construct a correlator of the form \eqref{E:Cdef}. As we now show, these are insufficient to construct a correlator of the form \eqref{E:Cdef}. Any rotational covariant version of $C$  must have a single epsilon tensor on $\mathbb{R}^{2+2k+4l}$ dotted into the $k$ momenta carried by the currents and $2l$ momenta carried by the energy-momentum insertions, leaving $2+k+2l$ antisymmetric indices. This tensor is orthogonal to all momenta. In addition $C$ has $2+4l$ free indices corresponding to the stress tensor insertions and $k$ free indices corresponding to the current insertions. The symmetry of the stress tensor $T^{\mu\nu} = T^{\nu\mu}$ implies that only $1+2l$ of the $2+4l$ indices may be antisymmetrized. Consequently, all $1+2l+k$ of the independent indices corresponding to the stress tensor and current insertions must appear in the epsilon tensor, leaving one index on the epsilon tensor and the remaining $1+2l$ symmetric indices on the stress tensors. However, this dangling index on the epsilon tensor cannot be one the remaining $1+2l$ symmetric indices (as then two indices on the same stress tensor insertion would appear in the epsilon tensor), nor can it be contracted with one of the momenta. Thus, we cannot write a rotationally covariant tensor with the correct symmetric properties of $C$. We conclude that in order for the partition function to consistently reproduce correlators in the Euclidean vacuum, we must have $\langle T_{cov}^{tr} \rangle = 0$. We will see that this provides non-trivial constraints on the coefficients of $W_{trans}$. These constraints will be captured by the replacement rule \eqref{E:RRv1}.

Before proceeding with the actual computation of $\langle T_{cov}^{tr} \rangle$ in the backgrounds of interest, let us pause and review our approach to proving the replacement rule \eqref{E:PT1V2} one more time. We work in a particular background \eqref{E:Ndef} which is specified by the numbers $k$, $l$ and $m_i$. In such a background we compute $\langle T^{tr}_{cov} \rangle$ and require it to vanish. As we will show, such a requirement will give us a constraint of the form \eqref{E:RRv1} with $\fP_{trans}$ given by an expression of the form \eqref{E:Ptransrep} which will be completely determined by $k$, $l$ and $m_i$. Varying over all possible values of $k$, $l$ and $m_i$ will allow us to probe all possible contributions to $\fP_{trans}$ thus proving the replacement rule \eqref{E:PT1V2} for arbitrary coefficients of the anomaly polynomial.

Before ending this Subsection, let us check that the number of configurations constructed from the backgrounds of the form \eqref{E:Ndef} with appropriate fluxes, is equal or greater to the number of coefficients in $\fP_{trans}$. Such a check is a straightforward exercise in counting. Consider a term $\fF_T^{2q} \wedge \form{p}(\fF,\fR)$ in $\fP_{trans}$ per~\eqref{E:Pcsb}, where $\form{p}$ is a product of various Chern classes of $\fF$ and Pontryagin classes of $\fR$.\footnote{We remind the reader that terms in $\fP_{trans}$ with an even number of $\fF_T$'s are CPT-preserving, while terms with an odd number are CPT-violating.} Suppose that this term has $2r$ powers of $\fR$, in which case it must have $s=n+1-2(q+r)$ powers of $\fF$. There are then $r$ independent Pontryagin classes that may be built out of this many powers of $\fR$, and $s$ independent Chern classes built out of this many powers of $\fF$.

Now let us count the number of backgrounds~\eqref{E:RRgauge} and~\eqref{E:RRmetric}. These backgrounds have $2k$ directions threaded by the flavor flux $\fF$ so that $\fF^k$ is nonzero, and so there are $k$ different Chern classes that may be built out of $\fF$ on this background. We also have $\sum_i 4m_i \equiv 4m = 2(n-1-k-2l)$ directions threaded by the Riemann curvature on the $\mathbb{CP}^{2m_i}$ directions; there are then $m$ different Pontryagin classes which may be built out of $\fR_{\mathbb{CP}}$. However, choosing $l \equiv q-1$ and $m\equiv r$, we see that there are $k=n+1-2(q+r) =s$ powers of $\fF$ in this background. Thus, there are as many independent terms with $2q$ factors of $\fF_T$ and $2r$ factors of $\fR$ in $\fP_{trans}$ as there are independent backgrounds of the form~\eqref{E:RRgauge} and~\eqref{E:RRmetric} with $l=q-1$ and $4r$ directions filled by the $\mathbb{CP}^{2m_i}$ spaces. Hence we have a one-to-one and onto map between the coefficients of $\fP_{trans}$ and the backgrounds~\eqref{E:RRgauge} and~\eqref{E:RRmetric}. Demanding $\langle T_{cov}^{tr}\rangle=0$ on all such backgrounds is sufficient to determine $\fP_{trans}$.

\subsection{Contributions to $\langle T_{cov}^{tr} \rangle$}
\label{S:perturbative}

In Sections \ref{S:intro} and \ref{S:HydroCS} we have advocated for a separation
\begin{equation}
	W_{QFT} = W_{non-anomalous}+W_{trans}+W_{anom}
\end{equation}
where $W_{anom}$ reproduces the anomalous variation of $W_{QFT}$, $W_{trans}$ corresponds to Chern-Simons terms on the base manifold and the remaining gauge and diffeomorphism invariant terms are collected in $W_{non-anomalous}$. In what follows we will argue that only $W_{trans}+W_{anom}$ may contribute to $\langle T_{cov}^{tr} \rangle$ for the backgrounds described in the previous Subsection. In other words, we argue that the variation of $W_{non-anomalous}$ with respect to small perturbations of the metric in the $tr$ directions vanishes. In the next Subsection we will study the variation of $W_{trans}+W_{anom}$ with respect to such perturbations and choose $W_{trans}$ such that $\langle T_{cov}^{tr} \rangle = 0$.

We start by enumerating the building blocks for all possible local tensors structures. In equilibrium the temperature, fluid velocity and chemical potential are local expressions of the background fields (see Equation \eqref{E:hydroVarDef}). For the backgrounds at hand, we find that
\beq
T = \frac{1}{2\pi r}\,, \qquad u^{\mu}\partial_{\mu} = \frac{1}{r}\partial_0\,, \qquad \mu = 0\,.
\eeq
First order gradients of these solutions satisfy
\beq
\label{E:gradientterms}
	D_{\mu}u_{\nu} = - u_{\mu}a_{\nu} + \omega_{\mu\nu}\,,
\qquad
(D_{\mu}+a_{\mu})T=0\,,
\eeq
with
\beq
\label{E:RRoneDeriv}
	a_{\mu}dx^{\mu} = \frac{dr}{r}\,, \qquad 
	2\fomega = -\frac{r}{2} f_{ij}dy^i \wedge dy^j\,, \qquad
	\form{F} = \frac{1}{2} B_{ab} dx^a \wedge dx^b\,,
\eeq
where in the last entry we have reproduced \eqref{E:RRF} for convenience. Here, the acceleration $a_{\mu}$ is given by $a_{\mu} = u^{\nu}D_{\nu}u_{\mu}$. Using \eqref{E:gradientterms} the spin chemical potential is given by
\beq
\label{E:muR}
(\mu_R)\ab{\mu}{\nu} = T D_{\nu}\left( \frac{u^{\mu}}{T}\right) = - (a^{\mu} u_{\nu}-u^{\mu}a_{\nu} + \omega\ab{\mu}{\nu})\,.
\eeq
Second order gradients are given by
\begin{align}
\begin{split}
\label{E:RRtwoDeriv}
D_{\mu}a_{\nu} &= - a^2 u_{\mu}u_{\nu} - a_{\mu}a_{\nu}\,, \qquad D_{\mu}(T \omega_{\nu\rho}) = 0\,,\qquad D_{\mu}F_{\nu\rho}=0\,,
\\
\omega_{\mu\nu}a^{\nu}&=0\,, \qquad \omega_{\mu\nu}\omega\ab{\nu}{\rho} = 0\,.
\end{split}
\end{align}
Since $a_{\mu}$ is along the $r$ direction, $\omega_{\mu\nu}$ and $a_{\mu}$ are covariantly constant in the $\{x^a,y^i,z^{\alpha}\}$ directions. In addition, the Riemann curvature tensor is given by
\begin{align}
\begin{split}
\label{E:RRriemann}
R\ab{\mu}{\nu\rho\sigma} = &-(\mu_R)\ab{\mu}{\nu}(2\omega_{\rho\sigma} ) + 2 \omega\ab{\mu}{\nu}(a_{\rho}u_{\sigma}-a_{\sigma}u_{\rho}) +(u^{\mu}a_{\sigma}-a^{\mu}u_{\sigma})\omega_{\nu\rho}- (u^{\mu}a_{\rho}-a^{\mu}u_{\rho})\omega_{\nu\sigma}
\\
& + \omega\ab{\mu}{\sigma}(u_{\nu}a_{\rho}-u_{\rho}a_{\nu}) - \omega\ab{\mu}{\rho}(u_{\nu}a_{\sigma}-u_{\sigma}a_{\nu})+\omega\ab{\mu}{\rho}\omega_{\nu\sigma} -\omega\ab{\mu}{\sigma}\omega_{\nu\rho} + (\RCP)\ab{\mu}{\nu\rho\sigma}\,,
\end{split}
\end{align}
where $(\RCP)^{\mu}{}_{\nu\rho\sigma}$ is the Riemann curvature of the $\mathbb{CP}$ spaces constructed from $G_{\alpha\beta}$. 

Since $(\RCP)^{\mu}{}_{\nu\rho\sigma}$ is covariantly constant it follows that, in the background we are considering, all tensor structures one can construct are given by contractions of $\{a_{\mu},\,\omega_{\mu\nu},\,F_{\mu\nu}\,,(\RCP)\ab{\mu}{\nu\rho\sigma}\}$, the velocity field $u_{\mu}$, the metric $g_{\mu\nu}$, and the epsilon tensor, $\epsilon^{\mu_1\hdots \mu_{2n}}$ (we take $\epsilon^{01\hdots 2n} = +1/\sqrt{-g}$) perhaps multiplied by some function of the temperature $T$. 

Consider the dependence of $\langle T_{cov}^{tr} \rangle$ on the $x$, $y$ and $z$ coordinates. Due to the product structure of our metric \eqref{E:RRmetric}, the component $T_{cov}^{tr}$ of the stress tensor behaves as a scalar with respect to the $\mathbb{CP}^{m_i}$ spaces. By symmetry, it must be independent of the $z^{\alpha}$ coordinates. Furthermore, we have seen that all tensor structures which we may use to construct $T^{\mu\nu}$ are independent of the $y^i$ and $x^a$ coordinates. Thus, $T_{cov}^{tr}$ must be independent of the $x$, $y$ and $z$ coordinates. Or, it must be equal to its average over the $y^i$, $x^a$ and $z^{\alpha}$ coordinates. Now, the average value of $T_{cov}^{tr}$ over the directions transverse to $t$ and $r$ can be obtained by varying $W_{cov}$ with respect to a metric perturbation $\delta g_{tr}(r)$. In equations
\begin{equation}
	\langle T_{cov}^{tr}  \rangle = \frac{2}{\sqrt{-g}} \frac{\delta W_{cov}}{\delta g_{tr}}\,.
\end{equation}

Let us consider the contribution of $W_{non-anomalous}$ to $\langle T_{cov}^{tr} \rangle$,
\begin{equation}
	\delta_{g_{tr}(r)} W_{non-anomalous}  = \int d^{2n}x \sqrt{-g} \frac{1}{2} \delta g_{tr}(r) T^{tr}_{non-anomalous}\,.
\end{equation}  
Perturbing the metric at  linear order in $\delta g_{tr}(r)$, is equivalent to an infinitesimal coordinate transformation of the metric~\eqref{E:RRmetric}. Being an infinitesimal coordinate transformation the covariant relations \eqref{E:RRtwoDeriv} and~\eqref{E:RRriemann} must still hold. Thus, we should be able to construct $\delta_{g_{tr}(r)} W_{non-anomalous}$ from $\{a_{\mu},\,\omega_{\mu\nu},\,F_{\mu\nu}\,,(\RCP)\ab{\mu}{\nu\rho\sigma}\}$, the velocity field $u_{\mu}$, the metric $g_{\mu\nu}$, and the epsilon tensor, $\epsilon^{\mu_1\hdots \mu_{2n}}$ evaluated on the background \eqref{E:RRmetric} perturbed by $g \to g+\delta g_{tr}(r)dtdr$. But the only tensor structures which are linear in $\delta g_{tr}$ are the metric $g_{\mu\nu}$ and the velocity field $u_{\mu}$. It is then straightforward  to check that there are no gauge-invariant scalars which are linear in $\delta g_{tr}$. This shows that no local term in $W_{non-anomalous}$ can contribute to the one-point function of $T_{cov}^{tr}$ in the 
background~\eqref{E:RRgauge} and~\eqref{E:RRmetric}. In the same way, there are no local terms in $W_{non-anomalous}$ which are non-analytic in derivatives, e.g. $\exp(-cT^2/a_{\mu}a^{\mu})$, which contribute to $\langle T_{cov}^{tr}\rangle$. It is somewhat subtle to argue that non-local terms in $W_{non-anomalous}$ do not contribute to $\langle T^{tr}_{cov}\rangle$ either. We refer the reader to~\cite{Jensen:2012kj} for further discussion.

\subsection{Constraining the thermal anomaly polynomial}
\label{S:RR}

Consistency of the Euclidean vacuum implies that $\langle T_{cov}^{tr} \rangle = 0$. We have argued that $\langle T_{cov}^{tr} \rangle$ can not receive contributions from $W_{non-anomalous}$. What remains is to compute the contribution of $W_{trans}+W_{anom}$ to $\langle T_{cov}^{tr} \rangle$.

By~\eqref{E:TT} we have
\beq
\label{E:T0r1}
\langle T^{tr}_{cov}\rangle = u^t q_T^r + u^r q_T^t + D_{\rho}\left[ \spp_T^{t[r\rho]} + \spp_T^{r[t\rho]} - \spp_T^{\rho(tr)}\right]\,.
\eeq 
Let us simplify this expression. Since $u^r = 0$ and $\spp_T$ is antisymmetric in its matrix-valued indices (the last two), the first and last terms in \eqref{E:T0r1} vanish. 

Both $\hodge \form{q}_T$ and $\hodge (\form{\spp}_T)\ab{\mu}{\nu}$ are $2n-1$ forms which are given by the velocity one form $\fu$ wedged with magnetic fields $\fB$, $\fB_R$ and the vorticity $\fomega$, multiplied by products of chemical potentials. The two forms $\fB$ and $\fomega$ have legs along the $\mathbb{R}^{2k}$ and $\mathbb{R}^{4l}$ planes respectively. The magnetic Riemann curvature follows from~\eqref{E:RRriemann} and is given by
\begin{align}
\begin{split}
\label{E:RRmagRiemann}
(B_R)\ab{\mu}{\nu\rho\sigma} = & - (\mu_R)\ab{\mu}{\nu}(2\omega_{\rho\sigma}) +(\omega\ab{\mu}{\sigma}u_{\nu}- u^{\mu}\omega_{\nu\sigma})a_{\rho} - (\omega\ab{\mu}{\rho}u_{\nu}-u^{\mu}\omega_{\nu\rho})a_{\sigma}
\\
& + \omega\ab{\mu}{\rho}\omega_{\nu\sigma} - \omega\ab{\mu}{\sigma}\omega_{\nu\rho} + (\RCP)\ab{\mu}{\nu\rho\sigma}\,.
\end{split}
\end{align}
Consequently, $(\fB_R)\ab{\mu}{\nu}$ is a sum of four types of forms: $\fomega$, $\fa \wedge \omega_{\mu\nu}dx^{\nu}$, $\omega_{\mu\rho}\omega_{\nu\sigma}dx^{\rho}\wedge dx^{\sigma}$, and $(\fRCP)\ab{\mu}{\nu}$, the curvature form on the $\mathbb{CP}$ spaces. 
Let us focus on the dependence of $\hodge \form{q}_T$ and $\hodge(\form{\spp}_T)\ab{\mu}{\nu}$ on the various form fields and ignore the contractions of the free indices for the time-being. The heat and spin current are given by
\beq
\label{E:RRparameterize}
\hodge \form{q}_T, \hodge (\form{\spp}_T) \sim \fu \wedge \fB^{q_1} \wedge \fomega^{q_2} \wedge (\fa \wedge \omega_{\mu\nu}dx^{\nu})^{q_3} \wedge (\omega_{\mu\nu}\omega_{\rho\sigma}dx^{\rho}dx^{\sigma})^{q_4} \wedge (\fRCP)^{q_5}\,,.
\eeq
Equation \eqref{E:RRparameterize} follows since $\hodge \form{q}_T$ and $\hodge \form{L}_T$ are $2n-1$ forms and the only form fields with legs in the $\mathbb{R}^{2k}$ plane and $\mathbb{CP}$ spaces are $\form{B}$ and $\form{\fRCP}$. Thus, $q_1=k$ and $q_5=2m$. Recall that  each factor of $\omega_{\mu\nu}$ can appear at most once in our perturbative counting and that $\form{u} = -r (dt+b_{ij}y^i dy^j/2)$ and $\form{a} = dr/r$. This implies that $q_3=1$ or $q_3=0$. In either case, $\form{u}$ must support the $dt$ direction and then $\hodge \form{q}_T$ and $\hodge \form{L}_T$ must fill in at least $4l-1$ indices in $\mathbb{R}^{4l}$. Therefore we must have $q_4=0$ and $q_2=2l$ and $q_3=0$ or $q_2=2l-1$ and $q_3=1$.
In the first case, the Hodge dual one-form is along the $r$ direction, and in the second along one of the $y^i$ directions. For each case, each nonzero $\omega_{\mu\nu}$ then appears exactly once, and so we can ignore the dependence of $\fu$, $(\mu_R)\ab{\mu}{\nu}$, and the Christoffel connection $\Gamma\ab{\mu}{\nu\rho}$ on $b_{ij}$. Functionally, this allows us to take
\begin{align}
\begin{split}
\label{E:simpleumrGCP}
	\fu &= - r dt + \mathcal{O}(b)\,, \\
	(\mu_R)\ab{\mu}{\nu} &= u^{\mu}a_{\nu} -a^{\mu}u_{\nu} + \mathcal{O}(b)\,, \\
	\Gamma\ab{\mu}{\nu\rho} &= - (\mu_R)\ab{\mu}{\nu}u_{\rho}-u^{\mu}u_{\nu}a_{\rho}+(\Gamma_{{}_\mathbb{CP}})\ab{\mu}{\nu\rho} + \mathcal{O}(b) \\
\end{split}
\end{align}
when deriving $\langle T_{cov}^{tr} \rangle$ from~\eqref{E:RRparameterize}. Here, $(\Gamma_{{}_\mathbb{CP}})\ab{\mu}{\nu\rho}$ is the Christoffel connection constructed from $G_{\alpha\beta}$. 

Since $\hodge (\form{\spp}_T)\ab{\mu}{\nu}$ has a leg along $dt$, it follows that $(\spp_T)\ab{t\mu}{\nu}= \mathcal{O}(b^{2l+1})$ in our perturbative scheme. Further, all tensor structures which comprise the spin current are covariantly constant in the $\{x^a,y^i,z^{\alpha}\}$ directions. 
Hence~\eqref{E:T0r1} simplifies to
\begin{align}
\begin{split}
\label{E:simpleTtr}
\langle T^{tr}_{cov}\rangle &= \frac{1}{r}q_T^r + D_r \spp_T^{rtr}+ D_0 \spp_T^{trt}
\\
& = \frac{1}{r}q_T^r + \left(\partial_r \spp_T^{rtr} + \Gamma\ab{r}{\mu r}\spp_T^{\mu tr} + \Gamma\ab{t}{\mu r} \spp_T^{r\mu r} + \Gamma\ab{r}{\mu r} \spp_T^{rt\mu} \right)
\\
 & \qquad + \left(\partial_t \spp_T^{trt} + \Gamma\ab{t}{\mu t} \spp_T^{\mu rt} + \Gamma\ab{r}{\mu t} \spp_T^{t\mu t} + \Gamma\ab{t}{\mu t} \spp_T^{tr\mu}\right)
\\
& = \frac{1}{r}q_T^r + \partial_r \spp_T^{rtr} \,,
\end{split}
\end{align}
where we have used that $\Gamma\ab{r}{\mu\nu}= \mathcal{O}(b^{})$ along with $\spp_T^{t\mu\nu}= \mathcal{O}(b^{2l+1})$ and $\spp_T^{\mu(\nu\rho)}=0$. The expression for $\langle T^{tr}_{cov} \rangle$ depends only on the $r$ component of $(\spp_T)\ab{\mu\nu}{\rho}$. Going back to \eqref{E:RRparameterize} we observe that only $q_2=2l,q_3=0$ will contribute to the expectation value we are interested in. Consequently, we may focus on the terms in the heat and spin currents which are of the form
\beq
\label{E:formqL}
\hodge\form{q}_T, \hodge (\form{\spp}_T) \sim \fu \wedge \fB^{k} \wedge \fomega^{2l}\wedge (\fR_{{}_{\mathbb{CP}}})^{2m}\,.
\eeq

Since all the factors of $b_{ij}$ have already been accounted for in \eqref{E:formqL} we may simplify our expressions for the spin chemical potential and the magnetic component of the Riemann tensor.
The magnetic component of the Riemann curvature~\eqref{E:RRmagRiemann} may be approximated by 
\beq
\label{E:BR}
(B_R)\ab{\mu}{\nu\rho\sigma} = - (\mu_R)\ab{\mu}{\nu}(2\omega_{\rho\sigma}) + (\RCP)\ab{\mu}{\nu\rho\sigma}\,.
\eeq 
In matrix form we have
\begin{equation}
\label{E:BRmatrix}
	(\form{B}_R)^{\mu}{}_{\nu} = \begin{pmatrix}
		\epsilon \form{\mathfrak{f}} & 0 & 0 & 0 \\
		0 & 0 & 0 & 0 \\
		0 & 0 & 0 & 0 \\
		0 & 0 & 0 &\fRCP 
		\end{pmatrix}
\end{equation}
where by \eqref{E:RRoneDeriv}
\begin{equation}
\label{E:omegaf}
	2\form{\omega} =-r \form{\mathfrak{f}},
\end{equation}	
$\epsilon$ is the $2\times 2$ antisymmetric tensor on $\mathbb{R}^{1,1}$ representing the $t$ and $r$ directions (i.e., $\epsilon^{tr} = 1/r$), and the remaining blocks represents the $x,\,y,\,z$ directions.
The matrix form of the spin chemical potential is given by
\beq
\label{E:muRmatrix}
	(\mu_R)\ab{\mu}{\nu} =   
	\begin{pmatrix}
		 \frac{1}{r}\epsilon & 0 & 0 & 0 \\
		0 & 0 & 0 & 0 \\
		0 & 0 & 0 & 0 \\
		0 & 0 & 0 & 0
		\end{pmatrix}
\eeq
implying that
\beq
\label{E:muRSquared}
	(\mu_R)\ab{\mu}{\rho} (\mu_R)\ab{\rho}{\nu} =	\begin{pmatrix}
		\frac{1}{r^2} I & 0 & 0 & 0 \\
		0 & 0 & 0 & 0 \\
		0 & 0 & 0 & 0 \\
		0 & 0 & 0 & 0
		\end{pmatrix}
\eeq
which is a projection onto the $t$, $r$ coordinates.
Using~\eqref{E:BR}, $\hat{\fB}_R$ is approximated by
\beq
(\hat{\fB}_R)\ab{\mu}{\nu} = (\fRCP)\ab{\mu}{\nu} = 
	\begin{pmatrix}
		0 & 0 & 0 & 0 \\
		0 & 0 & 0 & 0 \\
		0 & 0 & 0 & 0 \\
		0 & 0 & 0  &\fRCP 
	\end{pmatrix}\,,
\eeq 
which obeys
\beq
\label{E:muRorthogHBR}
(\mu_R)\ab{\mu}{\rho} (\hat{\fB}_R)\ab{\rho}{\nu} = (\hat{\fB}_R)\ab{\mu}{\rho}(\mu_R)\ab{\rho}{\nu} =0
\eeq
for all intents and purposes. 

We are now in a position to compute the contribution of $\form{V}_{\fP}$ to $\langle T^{tr}_{cov} \rangle$. From~\eqref{E:currentP} we have
\beq
\label{E:starpq}
\hodge \form{q}_{\fP} = \frac{\partial \VP}{\partial(2\fomega)} = \frac{\fu}{2\fomega}\wedge \left( \frac{\hat{\fP}-\fP}{2\fomega} - \frac{\partial \hat{\fP}}{\partial(2\fomega)}\right)\,.
\eeq
Recalling that $\form{\hat{B}} = \form{B} + 2 \form{\omega}\mu$ and $\form{\hat{B}}_R = \form{B}_R + 2 \form{\omega}\mu_R$ we may decompose the rightmost term in the parenthesis on the right hand side of \eqref{E:starpq}
\beq
\frac{\partial \hat{\fP}}{\partial(2\fomega)} = \mu \cdot \frac{\partial \hat{\fP}}{\partial \fB}+ (\mu_R)\ab{\mu}{\nu} \frac{\partial \hat{\fP}}{\partial(\fB_R)\ab{\mu}{\nu}}\,.
\eeq
In the background we are considering $\mu=0$. Additionally, $\hat{\fB}_R$ appears quadratically in $\hat{\fP}$, so that the indices of the derivative $\partial\hat{\fP}/\partial (\fB_R)\ab{\nu}{\mu}$ are carried by factors of $\hat{\fB}_R$, e.g.,
\beq
\label{E:dhatdBR}
\frac{\partial}{\partial (\fB_R)\ab{\mu}{\nu}} \text{tr}(\hat{\fB}_R^{2p}) = \frac{\partial}{\partial(\fB_R)\ab{\mu}{\nu}}\text{tr}(\fB_R+(2\fomega)\mu_R)^{2p} = p (\hat{\fB}_R^{2p-1})\ab{\nu}{\mu}\,.
\eeq
Since $\mu_R$ dotted into $\hat{\fB}_R$ vanishes by~\eqref{E:muRorthogHBR}, we find that for the purpose of computing $\langle T^{tr}_{cov} \rangle$, 
\beq
\frac{\partial \hat{\fP}}{\partial (2\fomega)}= 0\,,
\eeq
so that
\beq
\hodge \form{q}_{\fP} = -\frac{\VP}{2\fomega}\,.
\eeq

Similarly, we find from \eqref{E:currentP} that
\beq
\label{E:starLP}
\hodge (\form{\spp}_{\fP})\ab{t}{r} = \frac{\partial \VP}{\partial (\fB_R)\ab{r}{t}} = \frac{\fu}{2\fomega}\wedge \left( \frac{\partial \fP}{\partial (\fB_R)\ab{r}{t}}-\frac{\partial \hat{\fP}}{\partial (\hat{\fB}_R)\ab{r}{t}}\right)\,.
\eeq
The second term in the parenthesis on the right hand side of \eqref{E:starLP} vanishes by our discussion around~\eqref{E:dhatdBR}. The first term in the parenthesis has a functional form which one may determine explicitly, but is somewhat unenlightening. We only require its dependence on $r$. It is not hard to see that parametrically
\beq
\label{E:hodgeLPtr}
	\hodge (\form{\spp}_{\fP})\ab{t}{r} \sim \fu \wedge (\fB)^k \wedge (\mu_R)^t{}_r\,(\mu_R \fomega)^{2l} \wedge (\fRCP)^{2m}\,.
\eeq
In components \eqref{E:hodgeLPtr} becomes
\beq
\label{E:LPrtr}
	\spp_{\mathcal{P}}^{rtr} \sim \epsilon^{tr}u_t \,\epsilon^{a_1\hdots a_{2k}} B_{a_1a_2}\hdots B_{a_{2k-1}a_{2k}} \epsilon^{i_1\hdots i_{4l}}(\mu_R)^{tr}(\mu_R \omega_{i_1i_2})\hdots (\mu_R\omega_{i_{4l-1}i_{4l}})\,, 
\eeq
where we have used that the non vanishing traces of $\fRCP$ on the $\mathbb{CP}$ spaces are proportional to the volume form on the same. Recall that the epsilon tensor on the $tr$ directions is given by $\epsilon^{tr} = +1/r$ and the other epsilon tensors  in \eqref{E:LPrtr} are given by the Levi-Civitta symbol. 
Following \eqref{E:RRoneDeriv}, \eqref{E:simpleumrGCP} and~\eqref{E:muRSquared} we find that $u_t \sim r$, $(\mu_r \form{\omega})^2\sim r^0$ and $\mu_r^{tr} \sim r^{-2}$, so that
\beq
\label{E:sppMonomial}
\spp_{\mathcal{P}}^{rtr} \sim \mathcal{O}\left( \frac{1}{r^2}\right)\,.
\eeq
Thus, \eqref{E:simpleTtr} reduces to
\beq
\label{E:VPdoesntSatisfyIt}
\langle T_{\mathcal{P}}^{tr}\rangle = \frac{1}{r}\left( q_{\mathcal{P}}^r - 2 \spp_{\mathcal{P}}^{rtr}\right)\,.
\eeq
One may verify that~\eqref{E:VPdoesntSatisfyIt} is generically nonzero. For instance, consider a two-dimensional theory with $\fP = c_g \text{tr}(\fR^2)$. In Appendix~\ref{A:examples} we show (among other things) that the appropriate heat and spin currents are given by
\beq
q^{\mu}_{\mathcal{P}} = -c_g (\mu_R)\ab{\rho}{\sigma}(\mu_R)\ab{\sigma}{\rho} \epsilon^{\mu\nu}u_{\nu}\,, \qquad \spp_{\mathcal{P}}^{\mu\nu\rho} = - 2 c_g (\mu_R)^{\nu\rho}\epsilon^{\mu\sigma}u_{\sigma}\,,
\eeq
which in the present instance gives
\beq
q_{\mathcal{P}}^r = -\frac{2c_g}{r^2}\,, \qquad \spp_{\mathcal{P}}^{rtr} = -\frac{2c_g}{r^2}\,, \qquad \langle T_{\mathcal{P}}^{tr}\rangle = \frac{2c_g}{r^3}\,.
\eeq

So far we have computed the contribution of $\form{V}_\fP$ to $\langle T^{tr}_{cov} \rangle$ and have shown that it is non zero. In order to satisfy the consistency condition with the Euclidean vacuum we need that the contributions coming from $W_{trans}$ precisely cancel those of $\form{V}_\fP$. 
As advertised in Section \ref{S:HydroCS} we posit that the correct coefficients in $W_{trans}$ are captured by a thermal anomaly polynomial $\fP_T$ defined in \eqref{E:PT1V2} via the replacement rule
\begin{equation}
\label{E:Rrule}
	\fP_T = \fP\left(\form{F},\,\form{p}_k(\form{R}) \to \form{p}_k(\form{R}) - \left( \frac{\form{F}_T}{2\pi} \right)^2 \wedge \form{p}_{k-1}(\form{R}) \right)
\end{equation}
with $\form{F}_T$ a fictitious gauge field which is set to zero at the end of the computation and the $\form{p}_k(\fR)$ are the Pontryagin classes of $\fR$ (defined in~\eqref{E:defpk}). The combined contribution of $W_{trans}$ and $\VP$ on the covariant currents is given by varying $\form{V}_T$ in place of $\VP$ where
\begin{equation}
	\form{V}_T = \frac{\form{u}}{2\form{\omega}} \wedge \left(\fP_T - \hat{\fP}_T \right) ,
\end{equation}
and we set $\form{B}_T=0$ after varying $\form{V}_T$.

In what follows, we will find it  convenient to rewrite the replacement rule \eqref{E:Rrule} in 
terms of a fictitious Riemann curvature living in two dimensions higher. Thus, we write the thermal anomaly polynomial as\footnote{Note that the replacement $\form{R} \to \form{\mathcal{R}}$ may be interpreted as the curvature of a $(2n+2)\times (2n+2)$ matrix-valued one-form $ \begin{pmatrix} \fGamma & 0 \\ 0 & i E \fA_T\end{pmatrix}$.} 
\begin{equation}
\label{E:RruleV2}
	\fP_T = \fP\left(\form{F},\,\form{R} \to \form{\mathcal{R}}\right)
\end{equation}
where $\form{\mathcal{R}}$ is formally given by a two form valued  $(2n+2)\times(2n+2)$ matrix
\begin{equation}
	\form{\mathcal{R}}^{M}{}_{N} = \begin{pmatrix} \form{R} & 0 \\ 0 & i E \form{F}_T \end{pmatrix}\,,
	\qquad \hbox{with}\qquad
	E = \begin{pmatrix} 0 & 1 \\ -1 & 0 \end{pmatrix}\,.
\end{equation}
Here, the first entry of $\form{\mathcal{R}}$ collectively refers to the $2n\times 2n$ matrix $\fR\ab{\mu}{\nu}$, and the second entry to the two fictitious directions we have added. Indeed, since $\form{\mathcal{R}}$ is block diagonal, its eigenvalues are the eigenvalues of $\fR\ab{\mu}{\nu}$ as well as $\pm \fF_T$.\footnote{The imaginary factor in $\form{\mathcal{R}}$ may be puzzling at first sight. The important feature of $\form{\mathcal{R}}$ is that its eigenvalues are those of $\fR\ab{\mu}{\nu}$ and $\pm \fF_T$. We have simply chosen a representative $\form{\mathcal{R}}$ that accomplishes this and moreover is antisymmetric, in analogy with $\fR\ab{\mu}{\nu}$.} As a result, we have
\beq
\label{E:barTrace}
\text{tr}(\form{\mathcal{R}}^{2p-1}) = 0\,, \qquad \text{tr}(\form{\mathcal{R}}^{2p}) = \text{tr}(\fR^{2p}) + 2 \fF_T^{2p}\,.
\eeq
Alternatively, the Pontryagin classes $\form{p}_k$, defined through the formal infinite sum
\beq
\label{E:defpk}
\text{det}\left( \mathbb{I} + \frac{v \fR}{2\pi}\right) = \sum_{k=0}^{\infty} v^k \form{p}_k(\fR)\,,
\eeq
satisfy
\beq
\label{E:barPontryagin}
\form{p}_k(\form{\mathcal{R}}) = \form{p}_k(\fR)-\left( \frac{\fF_T}{2\pi}\right)^2 \wedge \form{p}_{k-1}(\fR)\,.
\eeq
Thus, equation~\eqref{E:RruleV2} is equivalent to~\eqref{E:Rrule} as advertised.

Next we note that since $\fP_T$ is quadratic in $\form{B}_T$ we may set
\begin{equation}
\label{E:BBhat}
	\form{B}_T = 0\,,\qquad
	\hat{\form{B}}_T = 2 \form{\omega}\mu_T = 4\pi T \form{\omega} = -\form{\mathfrak{f}}\,,
\end{equation}
Giving us the identity
\begin{equation}
	\form{u} \wedge \fP(\form{F},\form{R}) = \form{u}\wedge \fP_T(\form{F},\form{R},\form{F}_T) \Big|_{\form{B}_T=0}
\end{equation}
and, using \eqref{E:barTrace},
\begin{align}
\begin{split}
	\form{u} \wedge \fP\left(\hat{\form{F}},\,\text{tr}\left(\hat{\form{\mathcal{R}}}^{2p}\right)\right) &= \form{u} \wedge \fP\left(\hat{\form{F}},\,\text{tr}\left(\left(\form{B}_R + 2  \form{\omega}\mu_R\right)^{2p}\right) + 2 (\form{B}_T + 2 \form{\omega}\mu_T)\right)\Big|_{\form{B}_T=0} \\
	&= \form{u} \wedge \hat{\fP}_T\Big|_{\form{B}_T=0}.
\end{split}
\end{align}
Thus, $\form{V}_T$ takes the alternate form
\begin{equation}
\label{E:alternateVT0}
	\form{V}_T = \frac{\form{u}}{2\form{\omega}} \wedge \left(\fP\left(\form{F},\hbox{tr}\left(\form{R}^{2p}\right)\right) - \fP\left({\form{F}},\hbox{tr}\left(\hat{\form{\mathcal{R}}}^{2p}\right)\right) \right)\,
\end{equation}
where we have used $\hat{\form{B}} = \form{B}$ due to $\mu=0$ in the backgrounds under consideration. For brevity we shall suppress the explicit dependence of $\fP$ on the form fields $\form{F}$, $\form{R}$ and $\form{\mathcal{R}}$. Hence, we denote the first term in the parenthesis on the right hand side of \eqref{E:alternateVT0} simply by $\fP$ and the second term in the same parenthesis by $\widetilde{\fP}$. In this new notation \eqref{E:alternateVT0} becomes
\begin{equation}
\label{E:alternateVT}
	\form{V}_T = \frac{\form{u}}{2\form{\omega}} \wedge \left(\fP  - \widetilde{\fP} \right)\,.
\end{equation}

To proceed, we write down the magnetic component of $\form{\mathcal{R}}$, $\form{\mathcal{B}}_{\mathcal{R}}$, and its hatted cousin in our backgrounds. We do so in a matrix notation, in which the first four entries refer to the $\{t,r\}$ directions, followed by the $\{x^a,y^i,z^{\alpha}\}$ directions. The fifth entry corresponds to the two fictitious directions we have added. We then have 
\begin{equation}
\label{E:calBR}
	(\form{\mathcal{B}}_\mathcal{R})\ab{M}{N} = \begin{pmatrix} 
		\epsilon \form{\mathfrak{f}} & 0 & 0 & 0 & 0 \\
		0 & 0 & 0 & 0 & 0  \\
		0 & 0 & 0 & 0 & 0  \\
		0 & 0 & 0 & \fRCP & 0  \\
		0 & 0 & 0 & 0 & 0 
		\end{pmatrix}\,,
	\qquad
	(\hat{\form{\mathcal{B}}}_\mathcal{R})\ab{M}{N} = \begin{pmatrix} 
		0 & 0 & 0 & 0 & 0  \\
		0 & 0 & 0 & 0 & 0  \\
		0 & 0 & 0 & 0 & 0  \\
		0 & 0 & 0 & \fRCP & 0  \\
		0 & 0 & 0 & 0 & -i E \form{\mathfrak{f}} 
		\end{pmatrix}
\end{equation}
where we have used \eqref{E:BBhat}. Equation \eqref{E:calBR} together with \eqref{E:BR} imply that $\form{u} \wedge \hbox{tr}\left(\form{R}^{2p}\right) = \form{u} \wedge \hbox{tr}\left(\hat{\form{\mathcal{R}}}^{2p}\right)$ vanish on the background we are considering from which we conclude that $\form{V}_T$ vanishes on the backgrounds under consideration. Thus,
\begin{equation}
	\hodge \form{q}_T = -\frac{\form{u}}{2\form{\omega}} \wedge \frac{\partial \form{\widetilde{\fP}}}{\partial (2\form{\omega})} \\
		= -\frac{\form{u}}{2\form{\omega}} \wedge  \frac{\partial \form{\mathfrak{f}}}{\partial (2\form{\omega})} \frac{\partial \hat{\form{\mathcal{B}}}_{\mathcal{R}}}{\partial \form{\mathfrak{f}}} \cdot\frac{\partial}{\partial {\hat{\form{\mathcal{B}}}}_{\mathcal{R}}} \fP\left(\form{B},\hbox{tr}\left(\hat{\form{\mathcal{B}}}_{\mathcal{R}}^{2p}\right) \right)\,.
\end{equation}
Using \eqref{E:BRmatrix}, \eqref{E:muRmatrix} and \eqref{E:calBR}, we find that in our backgrounds
\begin{align}
\begin{split}
\label{E:hatcalBtoB}
	\frac{\partial \form{\mathfrak{f}}}{\partial (2\form{\omega})} \frac{\partial \hat{\form{\mathcal{B}}}_{\mathcal{R}}}{\partial \form{\mathfrak{f}}} \cdot
	\frac{\partial}{\partial {\hat{\form{\mathcal{B}}}}_{\mathcal{R}}} \fP\left(\form{B},\hbox{tr}\left(\hat{\form{\mathcal{B}}}_{\mathcal{R}}^{2p}\right) \right)
	&=
	-\mu_R \cdot
	\frac{\partial}{\partial \form{B}_R} \fP\left(\form{B},\hbox{tr}\left( \form{B}_R^{2p}\right) \right) \\
	&=-2 \frac{\partial \fP}{\partial (\form{B}_R)_{rt} } \,.
\end{split}
\end{align}
Thus,
\begin{equation}
	\hodge \form{q}_T = \frac{\form{u}}{\form{\omega}} \wedge \frac{\partial \fP}{\partial (\form{B}_R)_{rt} }\,.
\end{equation}

For the spin current a computation similar to the one above gives us \
\begin{align}
\begin{split}
	\hodge(\form{L_{T}})^{tr}  &= 
		\frac{\form{u}}{2\form{\omega}} \wedge \left(\frac{\partial \form{\mathcal{P}}}{\partial (\form{B}_R)_{rt}} - \frac{\partial \widetilde{\form{\mathcal{P}}}}{\partial (\form{\mathcal{B}}_{\mathcal{R}})} \cdot \frac{\partial \form{\mathcal{B}}_{\mathcal{R}} }{\partial  (\form{B}_R)_{rt}}  \right) \\
		&=\frac{\form{u}}{2\form{\omega}} \wedge \left(\frac{\partial \form{\mathcal{P}}}{\partial (\form{B}_R)_{rt}} - \frac{\partial \widetilde{\form{\mathcal{P}}}}{\partial (\form{\mathcal{B}}_{\mathcal{R}})_{rt}}  \right) \\
		&=\frac{\form{u}}{2\form{\omega}} \wedge  \frac{\partial \form{\mathcal{P}}}{\partial (\form{B}_R)_{rt}} \,.
\end{split}
\end{align}
Using the same argument that led to \eqref{E:sppMonomial} we get
\begin{equation}
	L_T^{trt} \sim \mathcal{O}\left(\frac{1}{r^2}\right)\,.
\end{equation}
Therefore, \eqref{E:simpleTtr} implies that 
\begin{equation}
	\langle T_{cov}^{tr} \rangle = \frac{1}{r} \left(q_T^{r} - 2 L_T^{rtr}\right) = 0
\end{equation}
as required. This proves that the replacement rule \eqref{E:PT1V2} is a sufficient condition for satisfying both the Wess-Zumino consistency condition and consistency with the Euclidean vacuum. Further, given our counting argument at the end of Subsection~\ref{S:observables}, the replacement rule is also the unique solution to these consistency conditions.

\acknowledgments

We would like to thank T.~Azeyanagi, Z.~Komargodski, P.~Kovtun, S.~Minwalla, G.~S.~Ng, M.~Rangamani, A.~Ritz, M.~J.~Rodriguez, and D.~Son for useful conversations and correspondence. RL would like to thank various colleagues at the Harvard Society of Fellows for interesting discussions. KJ also thanks the Perimeter Institute for Theoretical Physics and the Universiteit van Amsterdam for their hospitality while a portion of this work was completed. We are grateful to the organizers of \textbf{Holography and Applied String Theory} at the Banff International Research Station for their hospitality while this work was in progress. Part of this work was carried out during the \textbf{Relativistic hydrodynamics and the gauge gravity duality workshop} at the Technion. KJ was supported in part by NSCERC, Canada and by the National Science Foundation under grant PHY-0969739. RL was supported by the Harvard Society of Fellows through a junior fellowship. AY is a Landau fellow, supported in part by the Taub foundation as well as the ISF under grand number $495/11$, the BSF under grant number $2014350$ the European commission FP7, under IRG $908049$ and the GIF under grant number $1156/2011$.

\appendix
\section{Examples}
\label{A:examples}

In the main text, we have dealt with anomaly-induced transport in very general terms. In what follows we will descend from the heights of abstract differential geometry and come down to earth by working out the details of the transport coefficients for a few particular cases. In this Appendix we will collect some results for field theories in two, four, and for fun, ten dimensions.  Anomaly-induced transport in two~\cite{Dubovsky:2011sk,Jain:2012rh,Valle:2012em,Jensen:2012kj}. and four~\cite{Son:2009tf,Neiman:2010zi,Banerjee:2012iz,Jensen:2012jy,Jensen:2012kj,Jensen:2013kka} dimensions are interesting for phenomenological reasons and have been extensively studied in the literature. 

\subsection{Two-dimensional theories}
Consider a two-dimensional theory with a $U(1)$ global symmetry and anomaly polynomial
\beq
\label{E:anomP2d}
\fP = c_A \fF\wedge \fF + c_g \fR\ab{\mu}{\nu} \wedge \fR\ab{\nu}{\mu}\,.
\eeq
In these conventions, the anomaly coefficients corresponding to a left-moving Weyl fermion are $c_s = 1/(4\pi)$ and $c_g=1/(96\pi)$. The thermal anomaly polynomial obtained using the replacement rule~\eqref{E:PT1V2} is given by
\begin{equation}
\label{E:anomPT2d}
	\fP_T = \fP+ 2 c_g \form{F}_T \wedge \form{F}_T\,.
\end{equation}
According to our discussion in Section \ref{S:WtransCS}, the second term on the right hand side of \eqref{E:anomPT2d} manifests itself as a Chern-Simons term of the Euclidean generating functional. Denoting the Chern-Simons terms of the Kaluza-Klein reduction of the theory over the thermal circle as $W_{trans}$ and going to transverse gauge as in Section~\ref{S:HydroCS} we find that
\beq
W_{trans} =  \frac{8\pi^2  c_g}{\beta^2}\int  dt\wedge \form{\mathfrak{a}}\,.
\eeq

To compute the covariant currents we use~\eqref{E:anomPT2d} and~\eqref{E:VTdef} to obtain the master function
\begin{equation}
\form{V}_T =  -2\fu \wedge \left[ c_A \left( \mu\fB + \mu^2 \fomega\right) + c_g\text{tr} \left( \muR \fBR + \muR^2 \fomega\right)\right]    -4 c_g \fu \wedge \left( \mu_T \fB_T + \mu_T^2 \fomega\right)\,,
\end{equation}
where we have defined a trace over matrix-valued forms to be
\beq
\text{tr} \left( A_1 A_2\hdots A_m\right) = (A_1)\ab{\alpha_1}{\alpha_2} (A_2)\ab{\alpha_2}{\alpha_3}\hdots (A_m)\ab{\alpha_{m}}{\alpha_1}\,.
\eeq
The flavor, heat, and spin currents are given by~\eqref{E:Ttransport} 
\begin{align}
\begin{split}
\hodge \form{J}_T & = -  2 c_A \mu  \fu \,,
\\
\hodge \form{q}_T & = -\left( c_A \mu^2 + c_g \tr (\muR^2) +2 c_g \mu_T^2\right) \fu
\\
\hodge( \form{\spp}_T)\ab{\mu}{\nu} & = - 2 c_g (\muR)\ab{\mu}{\nu} \fu\,,
\end{split}
\end{align}
which in coordinates gives (upon using $\mu_T = 2 \pi T$)
\begin{align}
\begin{split}
\label{E:RRanomTrans2d}
J_T^{\mu} & = - 2 c_A \mu \epsilon^{\mu\nu}u_{\nu}\,,
\\
q_T^{\mu} &= - \epsilon^{\mu\nu}u_{\nu}\left( c_A \mu^2 + c_g (\mu_R)\ab{\alpha}{\beta}(\mu_R)\ab{\beta}{\alpha} + 8 \pi^2 c_g T^2\right)\,,
\\
(\spp_T)\ab{\mu\alpha}{\beta} & = - 2 c_g \epsilon^{\mu\nu}u_{\nu}(\mu_R)\ab{\alpha}{\beta}\,,
\end{split}
\end{align}
Let us decompose the anomaly-induced energy-momentum tensor
\beq
\label{E:TTintoQandSpin}
T_T^{\mu\nu} = u^{\mu} q_T^{\nu} + u^{\nu}q_T^{\mu} + D_{\rho}\left[ \spp_T^{\mu[\nu\rho]} + \spp_T^{\nu[\mu\rho]} - \spp_T^{\rho(\mu\nu)}\right]\,,
\eeq
into longitudinal and transverse components,
\beq
\label{E:TT2dDecomp}
T_T^{\mu\nu} = \mathcal{E}_T u^{\mu}u^{\nu} + \mathcal{P}_T P^{\mu\nu} + \mathcal{Q}\left( u^{\mu}\epsilon^{\nu\rho}u_{\rho} + u^{\nu}\epsilon^{\mu\rho}u_{\rho}\right)\,,
\eeq
where in two dimensions, a transverse vector is necessarily proportional to $\epsilon^{\mu\nu}u_{\nu}$, and there are no transverse traceless symmetric tensors. Plugging the anomaly-induced transport~\eqref{E:RRanomTrans2d} into~\eqref{E:TTintoQandSpin}, and using various differential identities that hold in equilibrium, we find that
\beq
\mathcal{E}_T = \mathcal{P}_T = 0\,, \qquad \mathcal{Q}_T = - (c_A \mu^2 + 8\pi^2 c_g T^2) + 2c_g \frac{D^{\mu}D_{\mu}T}{T}\,,
\eeq 
in accordance with the literature.

Previously, pure $U(1)$ anomalies in two-dimensional fluids have been considered from an effective action viewpoint in~\cite{Dubovsky:2011sk} and from the hydrostatic generating functional viewpoint in~\cite{Jain:2012rh,Valle:2012em,Banerjee:2012cr,Jensen:2012kj}. In~\cite{Jain:2012rh,Valle:2012em,Banerjee:2012cr}, the authors did not impose the replacement rule which results in several unfixed coefficients. One may reproduce the results of \cite{Jain:2012rh,Valle:2012em,Banerjee:2012cr} by using
\begin{equation}
	\fP_{trans} = \ \tilde{c}_1 \form{F}_T \wedge \form{F} + \tilde{c}_2 \form{F}_T^2
\end{equation}
with arbitrary $\tilde{c}_i$'s from which one obtains
\begin{align}
\begin{split}
\label{E:anomTrans2d}
J_T^{\mu} & = - \epsilon^{\mu\nu}u_{\nu} \left( 2 c_A \mu + 2\pi \tilde{c}_1 T \right)\,,
\\
q_T^{\mu} & = - \epsilon^{\mu\nu}u_{\nu} \left( c_A \mu^2 + c_g (\muR)\ab{\alpha}{\beta}(\muR)\ab{\beta}{\alpha}+ 2\pi \tilde{c}_1  \mu  T+4\pi^2 \tilde{c}_2T^2\right)\,,
\\
(\spp_T)\ab{\mu\alpha}{\beta} & = - 2 c_g \epsilon^{\mu\nu}u_{\nu} (\muR)\ab{\alpha}{\beta}\,,
\end{split}
\end{align}
The $\tilde{c}_i$'s first appeared in~\cite{Jain:2012rh} (although the physics of $\tilde{c}_1$ only appears in~\cite{Banerjee:2012cr}). Where these results overlap with ours, they agree. 
If we now suggestively use the out of equilibrium relations
\begin{align}
\begin{split}
(\mu_R)\ab{\mu}{\nu} & = - u_{\nu}a^{\mu} - \frac{u^{\mu} }{T} D_{\nu}T + P\ab{\mu}{\nu}\vartheta\,, \qquad \vartheta\equiv D_{\mu}u^{\mu}\,,
\\
(\mu_R)\ab{\mu}{\nu}(\mu_R)\ab{\nu}{\mu} & = - \frac{2}{T}a^{\mu}D_{\mu}T +\vartheta^2 + \frac{\dot{T}^2}{T^2}\,, \qquad \dot{T} = u^{\mu}D_{\mu} T\,.
\end{split}
\end{align}
we find that the anomaly-induced stress tensor is of the form~\eqref{E:TT2dDecomp} with
\begin{align}
\begin{split}
\label{E:TT2dUgly}
\mathcal{E}_T & = 2 c_g \epsilon^{\mu\nu}u_{\nu}\left( 2 a_{\nu}\vartheta - \frac{\dot{T}a_{\nu} + D_{\nu} \dot{T}}{T} + \frac{\dot{T}D_{\nu}T}{T^2}\right)\,,
\\
\mathcal{P}_T & = 2c_g \epsilon^{\mu\nu}u_{\nu}\left( - \dot{a}_{\nu} - D_{\nu}\vartheta  + \frac{\dot{T}a_{\nu}+D_{\nu}\dot{T}}{T} - \frac{\dot{T}D_{\nu}T}{T^2}\right)\,,
\\
\mathcal{Q}_T & = - \left( c_A \mu^2 + 8\pi^2 c_g T^2\right)   + c_g \left( - 2D_{\mu}a^{\mu} + 2a^2 + \vartheta^2 + \frac{2\dot{T}\vartheta }{T} - \frac{\dot{T}^2}{T^2}\right)\,.
\end{split}
\end{align}
We emphasize that \eqref{E:TT2dUgly} is merely suggestive of the role of anomalies in real-time transport. Whether or not the non-equilibrium terms in \eqref{E:TT2dUgly} are truly fixed by the anomaly is beyond the scope of the current work.

\subsection{Four-dimensional theories}

The anomaly polynomial for a four-dimensional theory with a $U(1)$ global symmetry is given by
\beq
\label{E:anomP4d}
\fP = c_A \fF^3 + c_m \fF \wedge \fR\ab{\mu}{\nu}\wedge \fR\ab{\nu}{\mu}\,.
\eeq
In these conventions, the anomalies for a theory of a left-moving Weyl fermion are $c_A = 1/(24\pi^2)$ and $c_m = 1/(192\pi^2)$. The thermal anomaly polynomial is given by~\eqref{E:PT1V2} to be 
\beq
\label{E:anomPT4d}
\fP_T = \fP + 2 c_m \fF_T^2 \wedge \fF \,.
\eeq
In a transverse gauge as in Section~\ref{S:HydroCS}, the Chern-Simons terms are then given by
\beq
W_{trans} = \frac{8\pi^2  c_m}{\beta^2}\int dt \wedge  \hat{\fA}\wedge d\form{\mathfrak{a}} \,.
\eeq
The master function $\form{V}_T$ is given by
\begin{align}
\begin{split}
	\form{V}_T =&  -\fu \wedge \left[ c_A \left( 3 \mu \fB^2 + 6 \mu^2 \fB \wedge \fomega + 4 \mu^3 \fomega^2\right)  + c_m \left( 2(\fB + 2\fomega\mu)\wedge \text{tr}(\mu_R \fB_R + \mu_R^2 \fomega) + \mu\, \text{tr}(\fB_R^2)\right)\right]
	\\
	&   - 2 c_m \fu \wedge \left[ 2 (\fB+ 2\fomega\mu)\wedge (\mu_T \fB_T+\mu_T^2\fomega) + \mu \fB_T^2\right] \,.
\end{split}
\end{align}
Taking derivatives of $\form{V}_T$, we obtain the flavor, heat, and spin currents by~\eqref{E:Ttransport} to be
\begin{align}
\begin{split}
\hodge \form{J}_T & = -(6c_A\mu + 2\tilde{c}_1 \mu_T)\fu\wedge \fB - 2 c_m (\mu_R)\ab{\mu}{\nu} \fu \wedge (\fB_R)\ab{\mu}{\nu}
\\
& \qquad \qquad  - (3c_A \mu^2 +c_m \text{tr}(\mu_R^2) + 2 c_m \mu_T^2)\fu\wedge(2 \fomega)\,,
\\
\hodge \form{q}_T &  = - \left[ 3c_A \mu^2 + c_m \text{tr}(\mu_R^2) + 2 c_m \mu_T^2\right]\fu \wedge \fB - 2  c_m \mu (\mu_R)\ab{\mu}{\nu}\fu\wedge (\fB_R)\ab{\nu}{\mu}
\\
& \qquad \qquad - 2\left[c_A \mu^3 + c_m \mu \text{tr}(\mu_R)^2  + 2 c_m \mu_T^2\mu \right]\fu\wedge (2\fomega)\,,
\\
\hodge (\form{\spp}_T)\ab{\mu}{\nu} & = - 2c_m (\mu_R)\ab{\mu}{\nu} \fu \wedge \fB -2 c_m \mu \fu\wedge\left( (\fB_R)\ab{\mu}{\nu}+(\mu_R)\ab{\mu}{\nu} (2\fomega) \right)\,.
\end{split}
\end{align}
In components these currents are
\begin{align}
\begin{split}
J_T^{\mu} & =- 6 c_A \mu B^{\mu} - 2 c_m (\mu_R)\ab{\alpha}{\beta}(B_R)\ab{\mu\beta}{\alpha} - \left( 3 c_A \mu^2 + c_m \text{tr}(\mu_R^2) + 8 \pi^2 c_m T^2\right)\omega^{\mu}\,,
\\
q_T^{\mu} & = - \left( 3 c_A \mu^2 + c_m \text{tr}(\mu_R^2) + 8 \pi^2 c_m T^2\right)B^{\mu} - 2c_m \mu (\mu_R)\ab{\alpha}{\beta}(B_R)\ab{\mu\beta}{\alpha}
\\
& \qquad \qquad - 2 \left(c_A\mu^3 + c_m \mu\, \text{tr}(\mu_R)^2 + 8 \pi^2 c_m \mu T^2\right)\omega^{\mu}\,,
\\
(\spp_T)\ab{\mu\alpha}{\beta} & = - c_m \left(2 (\mu_R)\ab{\alpha}{\beta} B^{\mu} +2\mu (B_R)\ab{\mu\alpha}{\beta} + (\mu_R)\ab{\alpha}{\beta} \omega^{\mu}\right)\,,
\end{split}
\end{align}
where in a slight abuse of notation we have defined
\beq
B^{\mu} = \half\epsilon^{\mu\nu\rho\sigma} u_{\nu}F_{\rho\sigma}\,, \qquad (B_R)\ab{\mu\alpha}{\beta} = \half \epsilon^{\mu\nu\rho\sigma} u_{\nu} R\ab{\alpha}{\beta\rho\sigma}\,, \qquad \omega^{\mu} = \epsilon^{\mu\nu\rho\sigma} u_{\nu} \partial_{\rho}u_{\sigma}\,.
\eeq

As in two dimensions, to make contact with the literature prior to \cite{Jensen:2012kj} and this work, we start with an extended polynomial which is not fixed by the replacement rule
\beq
\fP_{trans} =  \tilde{c}_1 \fF_T \wedge \fF^2 + \tilde{c}_2 \fF_T \wedge \fR\ab{\mu}{\nu} \wedge \fR\ab{\nu}{\mu} + \tilde{c}_3 \fF_T^2 \wedge \fF + \tilde{c}_4 \fF_T^3\,,
\eeq
where the $\tilde{c}_i$'s are not fixed. The $\tilde{c}_i$'s correspond to the Chern-Simons coefficients in the hydrostatic $W_{trans}$, which in transverse gauge is
\beq
W_{trans} = \int dt \wedge \left( -\frac{2\pi \tilde{c}_1}{\beta}\hat{A}\wedge d\hat{A} - \frac{2\pi \tilde{c}_2 }{\beta} \,\text{tr}\left( \hat{\form{\Gamma}}\wedge d\hat{\form{\Gamma}} + \frac{2}{3}\hat{\form{\Gamma}}^3\right)+\frac{4\pi^2 \tilde{c}_3}{\beta^2} \hat{\fA}\wedge d\form{\mathfrak{a}}  -\frac{8\pi^3\tilde{c}_4}{\beta^3} \form{\mathfrak{a}}\wedge d\form{\mathfrak{a}}\right)\,.
\eeq
Following our standard analysis we find that
\begin{align}
\begin{split}
\label{E:anomTrans4d}
J_T^{\mu} & = - (6 c_A \mu + 2\tilde{c}_1 \mu_T)B^{\mu} - 2 c_m (\mu_R)\ab{\alpha}{\beta} (B_R)\ab{\mu\beta}{\alpha} 
\\
& \qquad \qquad - \left(3c_A \mu^2 + c_m \text{tr}(\mu_R^2) + 2 \tilde{c}_1 \mu_T \mu + \tilde{c}_3 \mu_T^2\right)\omega^{\mu}\,,
\\
q_T^{\mu} & = -\left(3c_A\mu^3 + c_m \text{tr}(\muR)^2 + 2\tilde{c}_1 \mu_T \mu + \tilde{c}_3\mu_T^2\right)B^{\mu} - 2(c_m\mu+\tilde{c}_2\mu_T)(\mu_R)\ab{\alpha}{\beta}(B_R)\ab{\mu\beta}{\alpha}
\\
& \qquad \qquad - 2 \left( \tilde{c}_A \mu^3 + c_m \mu \text{tr}(\mu_R^2) + \tilde{c}_1 \mu_T \mu^2 + \tilde{c}_2 \mu_T \text{tr}(\mu_R^2) + \tilde{c}_3 \mu_T^2\mu + \tilde{c}_4\mu_T^3\right)\omega^{\mu}\,,
\\
(\spp_T)\ab{\mu\alpha}{\beta} &= - 2 c_m (\mu_R)\ab{\alpha}{\beta}B^{\mu} - 2(c_m \mu + \tilde{c}_2 \mu_T)(B_R)\ab{\mu\alpha}{\beta} - 2(c_m\mu + \tilde{c}_2 \mu_T)(\mu_R)\ab{\alpha}{\beta}\omega^{\mu}\,.
\end{split}
\end{align}
The coefficients $\tilde{c}_3$ and $\tilde{c}_4$ first appeared in~\cite{Neiman:2010zi}, while $\tilde{c}_1$ first appeared in~\cite{Banerjee:2012iz,Jensen:2012jy}. The results derived in those works of course agree with~\eqref{E:anomTrans4d}.

\subsection{Ten dimensional theories.}
For fun, we conclude with a study of ten-dimensional theories. The anomaly polynomial for a ten dimensional theory with pure gravitational anomalies is given by\footnote{Since we consider a theory with gravitational anomalies alone, this example is for fun rather than phenomenology. For instance, it is difficult to see how it can be embedded in string theory.}
\begin{equation}
	\fP = c_1 \text{tr}(\form{R}^2)^3+c_2 \text{tr}(\form{R}^4)\text{tr}(\form{R}^2) + c_3 \text{tr}(\form{R}^6)\,,
\end{equation}
from which we obtain $\fP_T$ via the replacement rule~\eqref{E:PT1V2}
\begin{equation}
	\fP_T = \fP + 2 (4 c_1+2 c_2+c_3) \form{F}_T^6 + 2 (6 c_1 + c_2) \form{F}_T^4 \wedge  \text{tr}(\form{R}^2) + \form{F}_T^2 \wedge \left(6 c_1 \text{tr}(\form{R}^2)^2 + 2 c_2 \text{tr}(\form{R}^4) \right)\,.
\end{equation}
The master function is quite long, we give here only the leading contribution in a derivative expansion 
\begin{equation}
	\form{V}_T = - 2(4 c_1 +2 c_2+ c_3) (2\pi T)^6  \form{u} \wedge (2\form{\omega})^5 + \mathcal{O}(\form{B}_T^2) + \mathcal{O}(\partial^8)\,.
\end{equation}
This gives us
\begin{align}
\begin{split}
	\hodge\form{q}_T &= -10(4 c_1+2 c_2 + c_3) (2\pi T)^6 \form{u} \wedge (2\form{\omega})^4 \\
\end{split}
\end{align}
In components this gives
\begin{align}
\begin{split}
	q_T^{\mu} &= -\frac{10(4 c_1+2 c_2 + c_3)}{4!} (2\pi T)^6 \epsilon^{\mu\nu_1\ldots\nu_{9}}u_{\nu_1}\omega_{\nu_2\nu_3}\ldots \omega_{\nu_8\nu_9}\,.
\end{split}
\end{align}

\bibliographystyle{JHEP}
\bibliography{refs2}

\end{document}